\documentclass[aps,reprint,nofootinbib,nobibnotes,notitlepage,superscriptaddress,onecolumn,prd,
preprintnumbers,
 amsmath,amssymb
]{revtex4-2}

\usepackage[caption=false]{subfig}
\usepackage{braket}
\usepackage{float}
\usepackage{lipsum}
\usepackage{graphicx}
\usepackage{dcolumn}
\usepackage{bm}
\usepackage{natbib}
\usepackage{hyperref}
\hypersetup{
	colorlinks = true,
    linkcolor = Red,
    urlcolor  = Red,
    citecolor = Red
}
\usepackage[skip=4pt plus1pt, indent=10pt]{parskip}
\usepackage{microtype}

\usepackage{verbatim}
\usepackage[amssymb]{SIunits}
\usepackage{tabularx}
\usepackage[dvipsnames]{xcolor}
\usepackage{wasysym}

\usepackage{tikz}
\usepackage[compat=1.1.0]{tikz-feynman}

\def\be{\begin{equation}}
\def\ee{\end{equation}}

\usepackage{booktabs}

\usepackage{color}
\definecolor{darkgreen}{RGB}{0,120,0}

\definecolor{darkgreen}{RGB}{0,120,0}

\definecolor{darkgreen}{RGB}{0,120,0}
\newcommand{\resub}[1]{{\color{black}{#1}}}
\newcommand{\resubb}[1]{{\color{black}{#1}}}

\newcommand{\eV}{{\,\rm eV}}
\newcommand{\Mpc}{\text{Mpc}}

\newcommand{\hMpc}{h\,\mathrm{Mpc}^{-1}}

\newcommand*{\polybin}{\selectfont\textsc{PolyBin3D}\xspace}

\newcommand{\ld}{\Lambda{\rm CDM}}

\newcommand{\wa}{w_0w_a{\rm CDM}}
\newcommand{\bao}{{\rm BAO}}
\newcommand{\cmb}{{\rm CMB}}
\newcommand{\cmbpr}{\text{CMB-PR4}}
\newcommand{\sn}{{\rm SN}}
\newcommand{\Omk}{\Omega_{\rm K}}
\newcommand{\OmK}{o\Lambda{\rm CDM}}

\DeclareSymbolFont{toneletters}{T1}{\familydefault}{m}{it}
\SetSymbolFont{toneletters}{bold}{T1}{\familydefault}{bx}{it}

\DeclareMathSymbol\edth{\mathord}{toneletters}{"F0}

\usepackage{adjustbox}
\usepackage{dashbox}

\def\beq{\begin{eqnarray}}
\def\eeq{\end{eqnarray}}

\usepackage{empheq}

\def\aap{\rm{A\&A}}

\defcitealias{desi1}{Paper 1}
\defcitealias{desi3}{Paper 3}

\usepackage{xspace}
\newcommand{\paperone}{\citetalias{desi1}\xspace}

\begin{document}

\preprint{MIT-CTP/5960}

\title{{\Large Reanalyzing DESI DR1:}\\
2. Constraints on Dark Energy, Spatial Curvature, and Neutrino Masses}

\author{Anton~Chudaykin}
\email{anton.chudaykin@unige.ch}
\affiliation{D\'epartement de Physique Th\'eorique and Center for Astroparticle Physics,\\
Universit\'e de Gen\`eve, 24 quai Ernest  Ansermet, 1211 Gen\`eve 4, Switzerland}
\author{Mikhail M.~Ivanov}
\email{ivanov99@mit.edu}
\affiliation{Center for Theoretical Physics -- a Leinweber Institute, Massachusetts Institute of Technology, 
Cambridge, MA 02139, USA} 
 \affiliation{The NSF AI Institute for Artificial Intelligence and Fundamental Interactions, Cambridge, MA 02139, USA}
\author{Oliver~H.\,E.~Philcox}
\email{ohep2@cantab.ac.uk}
\affiliation{Leinweber Institute for Theoretical Physics at Stanford, 382 Via Pueblo, Stanford, CA 94305, USA}
\affiliation{Kavli Institute for Particle Astrophysics and Cosmology, 382 Via Pueblo, Stanford, CA 94305, USA}

\begin{abstract} 
    \noindent We carry out an independent re-analysis of the Dark Energy Spectroscopic Instrument (DESI) public dataset, focusing on extensions to the standard cosmological model, $\Lambda$CDM. 
    Utilizing the dataset and Effective Field Theory (EFT)-based pipeline described in \paperone \citep{desi1}, we constrain cosmological models with massive neutrinos ($\Lambda$CDM+$M_\nu$), spatial curvature ($o\Lambda$CDM), dynamical dark energy ($w_0w_a$CDM), and their combinations using the power spectrum and bispectrum of DESI galaxies and quasars. 
    Our work also presents the first measurements of relevant non-minimal cosmological parameters from the combination of cosmic microwave background (CMB) and DESI full-shape (FS) data, which are made possible thanks to carefully chosen priors on EFT parameters.
    We find that the addition the FS likelihood to DESI's baryon acoustic oscillation (BAO) data improves the limits on the spatial curvature by a factor of two over the BAO only results, though the improvements are less significant with the CMB data. 
The dark energy equation of state
figure-of-merit 
increases both
with and without the supernovae data (SNe),
by $\approx 30\%$ and $\approx 20\%$ relative to the CMB+BAO and CMB+BAO+SNe results, respectively. 
Our FS likelihood also yields the strongest CMB-independent  
constraint on the total neutrino mass $M_\nu<0.32~\eV$,
    with the $30\%$ improvement due to the bispectrum.
   In combination with the CMB, we find a $14\%$ improvement assuming the $\Lambda$CDM+$M_\nu$ model (yielding $M_\nu<0.059~\eV$), but this increases to $22\%$
   when using non-minimal backgrounds: $M_\nu<0.097~\eV$
   in $o\Lambda$CDM+$M_\nu$ and $M_\nu<0.13~\eV$ in $w_0w_a$CDM+$M_\nu$.   
Overall, our work illustrates that robust and substantial gains in constraining power
can be obtained by incorporating the FS power spectrum and bispectrum measurements in analyses of non-minimal cosmological models.
\end{abstract}

\maketitle

\section{Introduction}

\setlength{\parskip}{2pt plus1pt}

\noindent 
The standard cosmological paradigm, $\Lambda$CDM, comprises two unknown sectors: cold dark matter (CDM) and the cosmological constant ($\Lambda$). This provides a remarkably successful model for the large-scale Universe, capable of describing a wide range of observations with just six parameters: the densities of baryons and cold dark matter, the present expansion rate, the reionization optical depth, and the spectral properties of the inflationary perturbations. Despite its successes, the model faces important challenges. One such difficulty is theoretical -- the paradigm offers no explanation for the physical nature of dark matter nor the amplitude of the cosmological constant. Moreover, recent tests using high-precision cosmological datasets may indicate cracks in the model foundations, possibly hinting at a more nuanced dark energy sector \citep[e.g.,][]{DESI:2025zgx,DESI:2025fii,DESI:2025wyn}. In addition, the vanilla $\Lambda$CDM model assumes a specific value for the sum of neutrino masses, $\sum m_\nu =0.06\,\mathrm{eV}$,  which in principle, should be treated as an unknown parameter.

Cosmological predictions can be loosely divided into two regimes: background and perturbations. These are sensitive to different model components, with the former set primarily by the Universe's composition and Hubble parameter (which acts as a normalization constant), and the latter adding dependence on interactions and the gravitational sector (and thus physics beyond General Relativity). To probe the background, we can combine distance measurements across redshift, including those from Type Ia supernovae (SNe), galaxy surveys, and the Cosmic Microwave Background (CMB), which respectively calibrate the expansion history at low-, medium- and high-redshifts.

The first data release (DR1) of the Dark Energy Spectroscopic Instrument (DESI) has provided new precision measurements of the Universe's expansion history, utilizing the `Baryon Acoustic Oscillation' feature measured in a range of galaxy samples spanning redshifts $z\in[0.1,2.1]$ \citep{DESI:2024aax,DESI:2024uvr,DESI:2024lzq,DESI:2024mwx}. Interpreting these data in the context of the $\Lambda$CDM model reveals a slight discrepancy with the CMB predictions, notably favoring higher values of $H_0$ and lower values of $\Omega_m$ \citep{DESI:2024mwx}, with the latter disagreement increased when comparing to predictions from SNe \citep{Brout:2022vxf,DES:2024jxu,Rubin:2023ovl}. Moreover, the combined CMB+DESI dataset finds a weak preference for dynamical dark energy (DDE); adopting the linear `Chevallier-Polarski-Linder' (CPL) parametrization with equation of state $w(a) = w_0 + (1-a)w_a$, the data finds $w_0 > -1$ and $w_a < 0$ at $2.6\sigma$. Whilst these results are not statistically significant on their own, they become stronger in combination with SNe datasets \citep{Brout:2022vxf,DES:2025sig,Rubin:2023ovl,DES:2024jxu} (several of which have previously reported hints for non-standard dark energy).

The most recent DESI release (DR2) features two additional years of data, greatly increasing the precision of cosmological distance measurements \citep{DESI:2025zpo,DESI:2025zgx}. Whilst DR2 constraints on the Universe's perturbation sector are not yet available, the BAO results are broadly consistent with the predictions of $\Lambda$CDM (at $2.3\sigma$) \citep[e.g.,][]{Efstathiou:2025tie}, and find no evidence for a fixed, but non-standard, dark energy equation of state when combining with CMB and SNe datasets. Assuming the CPL parametrization, the joint analysis again prefers $w_0>-1, w_a<0$, with significances ranging from $2.8-4.2\sigma$, depending on the data combination.~\footnote{Using the updated DES-SN5YR data release~\cite{DES:2025sig} reduces the evidence for non-standard dark energy from $4.2\sigma$ to $3.2\sigma$ in the joint analysis.} Remarkably, DR2 is the first dataset which shows preference for dynamical dark energy even in the absence of SNe.

Whilst these results are exciting, they remain somewhat inconclusive. 
First, as highlighted in \citep{Cortes:2024lgw,Cortes:2025joz}, the data suffer from a coincidence problem with no preference for $w(a_{\rm eff})\neq -1$ at the effective scale factor of the observations $a_{\rm eff}\approx 0.75$.\footnote{This is equivalent to the observation that the $w_0,w_a$ contour is curiously aligned towards the cosmological constant prediction.} 
Second, although results based on different SN compilations are broadly consistent, supernova distance measurements remain subject to several systematics, including uncertainties in redshift calibration, modeling of foregrounds, and environmental effects. These systematics can potentially propagate into a spurious preference for evolving dark energy~\cite{Dhawan:2024gqy,Gialamas:2024lyw}. It is therefore crucial to ensure that supernova systematics are well understood and rigorously controlled.
Third, the conclusion on the preference for 
DDE diminishes even in the presence of SNe if the full-shape (FS) clustering information from the BOSS survey is added~\cite{Chen:2024vuf,Chen:2025jnr} (see however~\cite{Silva:2025twg}). 
Fourth, several works have discussed the differences between the low-redshift samples of SDSS BOSS and DESI \citep[e.g.,][]{Ghosh:2024kyd,Chudaykin:2024gol,Chaudhary:2025pcc}. Fifth, the CPL parametrization has also been extensively interrogated: though it is capable of describing a variety of dark energy models \citep{dePutter:2008wt}, it is important to note that the parameters are purely effective values, and do not fully encapsulate the dark energy expansion history \citep{Wolf:2023uno}.
Sixth, the role of higher-order terms in the $w(a)$ expansion is debated \citep{Nesseris:2025lke}, though \citep{DESI:2025fii} found that adding such terms does not significantly improve the fit to the combined data and would thus be disfavored from a model comparison perspective.
Finally, the evidence for dynamical dark energy weakens significantly in some analysis variants, e.g.~\cite{Huang:2025som,Efstathiou:2025tie}.

From a theoretical perspective, the best-fit dark energy model is puzzling. The data favors `phantom-crossing' dark energy, evolving from phantom ($w<-1$) at early times, to quintessent ($w>-1$) today. These preferences appear in both the CPL parametrization and in model-agnostic reconstructions \citep[e.g.,][]{Wolf:2025jlc,DESI:2025fii,DESI:2025wyn,Berti:2025phi,DESI:2025wzd,Chen:2025jnr}. Na\"ively, this implies a violation of the null energy condition (assuming a single scalar field model), which represents a serious theoretical challenge (involving a number of undesirable effects such as ghost instabilities) \citep{Caldwell:1999ew}.
The `phantom-crossing' is crucial to achieving a good fit to data, as physical models without this property (such as scalar-field quintessence) do not significantly improve the fit over the cosmological constant~\cite{Toomey:2025xyo}. In addition, a phantom crossing at $z\sim 1$ is observed in many different dark energy parameterizations~\cite{DESI:2025fii}.
That said, one can obtain an effective `phantom-crossing' in the presence of non-gravitational interactions between dark energy and other sectors; this may indicate the need for non-minimal coupling, oscillations, or interactions between dark energy and dark matter \citep[e.g.,][]{Das:2005yj,Carroll:2004hc,Brax:2023qyp,Khoury:2025txd,Silva:2025hxw,Wolf:2024eph,Wolf:2024stt,Bottaro:2024pcb,Wolf:2025jed,DESI:2025fii,Goldstein:2025epp,Smith:2024ibv,Burgess:2025vxs,Bedroya:2025fwh,Caldwell:2025inn,SanchezLopez:2025uzw,Li:2024qso,Gialamas:2025pwv}.  

Motivated by the above difficulties, it is important to consider alternative explanations for the anomalous growth history reported by DESI. A notable case is cosmic curvature; as discussed in \citep{DESI:2024mwx,DESI:2025zgx,Chen:2025mlf}, the combined CMB+BAO dataset prefers negative curvature ($\Omega_k>0$) at $2.1\sigma$. This parameter cannot be appreciably measured using SNe, and, if confirmed to be non-zero, would place intriguing constraints on inflationary physics. A wide variety of other models have been proposed: these include modified gravity \citep{Chudaykin:2024gol,Ishak:2024jhs,Yang:2024kdo,Chudaykin:2025gdn}, cosmic topology \citep{Philcox:2025faf}, non-standard neutrino physics \citep{Craig:2024tky,Green:2024xbb,DESI:2025ejh}, dark sector physics \citep{Graham:2025fdt}, early dark energy \citep{Chaussidon:2025npr}, and modifications to the optical depth \citep{Sailer:2025lxj,Allali:2025yvp}. Many of these scenarios also affect the Universe's perturbation sector; this will help to constrain and distinguish such models.

Another intriguing feature of the recent DESI analyses concerns neutrinos, which impact cosmological evolution both at the background and fluctuation level. Oscillation experiments provide information about the mass difference (splitting) between the neutrino states, suggesting a lower bound of $\sum m_\nu=0.06\,\mathrm{eV}$ (as adopted in the base $\Lambda$CDM model). Furthermore, they suggest that two massive states are nearly degenerate, giving rise to two options: the normal hierarchy 
(where the nearly degenerate states are low-lying), and the inverted hierarchy (where they are the most massive states). The neutrino masses are also unknown parameters of the Standard Model of Particle Physics, entering as Wilson coefficients of the dimension-5 Weinberg effective operators.\footnote{Here we adopt the modern view that the Standard Model is an effective field theory, which provides a simple and natural explanation to the parametrically small mass of neutrinos (compared to the masses of other leptons) from power-counting arguments.} The values of neutrino masses can thus provide information about the ultraviolet-completion of the Standard Model, which may have important consequences on fundamental physics in general.

Currently, the cosmological upper bounds on neutrino masses are much stronger than those obtained from particle physics experiments, though they can only constrain the total mass of neutrinos, $\sum m_\nu$ \citep{Archidiacono:2020dvx}. The current nominal constraints from CMB and DESI BAO are quite strong,  $M_\nu<0.06-0.08\eV$ depending on the CMB data~\cite{DESI:2025zgx}, and going down to $\sum m_\nu<0.05\,\mathrm{eV}$ at $95\%$CL when using the profile likelihood method~\cite{DESI:2025zgx}.
These strong limits are further supported by other CMB and large-scale structure datasets~\cite{Palanque-Delabrouille:2019iyz,Ivanov:2019hqk,Garny:2020rom,Kumar:2022vee,ACT:2023kun,Ivanov:2024jtl}.
Taken at face value, this disfavors the inverted hierarchy, thereby implying an important constraint on particle physics models. The major caveat behind these limits is that they strongly depend on the assumptions about the underlying cosmological model; for example, the bound is significantly loosened by the presence of dynamical dark energy or non-zero spatial curvature, allowing for the inverted hierarchy.\footnote{\resub{An interesting exception arises when non-phantom behavior, $w(z)\geq -1$, is imposed within the CPL parameterization. In this case, the $M_\nu$ bound is slightly tighter than the $\ld$ limit which can be attributed to the degeneracy between the dark energy equation of state, $w(z)$, and $M_\nu$~\cite{Vagnozzi:2018jhn}.}} 
\resub{Moreover, the $\ld$ posterior distributions are modulated by ``prior weight effects'', which arise when the prior on a parameter pulls the posterior away from its maximum likelihood value~\cite{Elbers:2025vlz}.
More general dark energy parametrizations can remove unsatisfactory prior weight effects, relaxing the tension with neutrino oscillation experiments.}
It is therefore natural to analyze neutrino masses in combination with these beyond-$\Lambda$CDM phenomena.

To resolve the DESI discrepancies, more data is needed. A promising avenue is to incorporate information from the full-shape of the galaxy power spectrum and bispectrum; these depend on both the background and perturbation sectors, facilitating precise constraints on the above model extensions. In this work, we place novel constraints on dark energy, curvature, and neutrinos, utilizing the FS clustering data from the DESI DR1 public release \citep{DESI:2025fxa,DESI:2024aax,DESI:2024jis,DESI:2024hhd}, in combination with theoretical models derived from the Effective Field Theory of Large Scale Structure (EFT, \citep[e.g.,][]{Ivanov:2019pdj,DAmico:2019fhj,Chen:2021wdi,Vlah:2015zda,Pajer:2013jj,Carrasco:2012cv,Mercolli:2013bsa,Porto:2013qua,Vlah:2015sea,Senatore:2014via,Senatore:2014vja,Senatore:2014eva,Angulo:2015eqa,Senatore:2017hyk,Baumann:2010tm,Assassi:2014fva,Assassi:2015jqa,Chen:2020zjt,Blas:2015qsi,Blas:2016sfa,Ivanov:2018gjr,McDonald:2006mx,McDonald:2009dh,Ivanov:2022mrd})\footnote{For the use of the EFT in full-shape
analyses see \citep{Ivanov:2019pdj,DAmico:2019fhj,Chen:2021wdi,Ibanez:2024uua,DAmico:2022osl,DAmico:2022gki,Ivanov:2024hgq,Ivanov:2021fbu,Ivanov:2021zmi,Chudaykin:2022nru,Ivanov:2021kcd,Ivanov:2023qzb,Cabass:2024wob,Zhang:2021yna,Chen:2022jzq,Cabass:2022ymb,Chen:2021wdi,Philcox:2021kcw,Holm:2023laa,Wadekar:2020hax,Colas:2019ret,Ivanov:2019hqk,Chen:2024vuf,He:2023oke,He:2023dbn,Xu:2021rwg,Chudaykin:2022nru,Chudaykin:2020ghx,Spaar:2023his,Lu:2025gki,Lu:2025sjg,DAmico:2020tty}, as well as the public code papers~\citep{McEwen:2016fjn,Fang:2016wcf,classpt,DAmico:2020kxu,Chen:2020fxs,Linde:2024uzr,Noriega:2022nhf}.}. This builds on the $\ld$ analysis pipeline presented in \citep{desi1} (hereafter \paperone), and extends the official DESI analysis of \citep{DESI:2024hhd} by adding the power spectrum hexadecapole moment (whose utility was demonstrated in \citep{Chudaykin:2020ghx}) and the large-scale bispectrum monopole (which has been analyzed only in the compressed ShapeFit paradigm \citep{NovellMasot:2025fju}). This is made possible through the quasi-optimal estimators and analysis pipeline discussed in \paperone (based on \citep{Philcox:2024rqr}), which allow for robust treatment of various systematic effects, including fiber collisions in both the power spectrum and bispectrum. 
We additionally adopt a new set of priors on EFT parameters, which allows us to obtain the first dark energy measurements from the CMB and combined DESI (BAO+FS) data without relying on supernova distance information.
By combining our DESI likelihood with CMB data, BAO from DESI DR2, and different compilations of SNe, we can place tight constraints on model extensions, both elucidating the impact of FS data and, where possible, obtaining constraints independent from the CMB or BAO.

The remainder of this paper is as follows. \S\ref{sec:data} describes the datasets used in this work, comprising the DESI DR1 FS measurements from \paperone (built upon data from \citep{DESI:2025fxa}), the DR2 BAO, the CMB, and Type Ia supernovae, as well as the theoretical model used herein. Our main results are presented in \S\ref{sec:results}: constraints on spatial curvature, time-varying dark energy, and neutrino masses (in conjunction with the above extensions). We conclude in \S\ref{sec:conclusions} with a discussion of implications and directions for future work. The appendices contain detailed studies of the effects of EFT parameter priors on the inference of cosmological parameters in the non-minimal models. 

\section{Data \& Theory}\label{sec:data}
\subsection{Full-Shape Data}\label{sec:data1}
\noindent In this work, we utilize the DESI DR1 power spectrum and bispectrum dataset measured from the public data release \citep{DESI:2025fxa}. This is described in detail in \paperone, and we summarize the salient points below. The dataset comprises six non-overlapping data chunks -- BGS, LRG1, LRG2, LRG3, ELG2, and QSO -- spanning a combined redshift range of $0.1\leq z\leq 2.1$ \citep{DESI:2024aax}. For each chunk, we combine data from the North and South galactic caps according to the respective survey areas. As discussed in \paperone, we measure the two- and three-point statistics from each chunk using quasi-optimal `unwindowed' estimators, as implemented in the \polybin code \citep{polybin3d,Philcox:2024rqr}. These are closely related to the usual windowed estimators but feature approximate mask-deconvolution, analogous to the CMB pseudo-$C_\ell$ approach. 

Utilizing an FFT-based algorithm, we jointly compute the summary statistics and binning matrices (which account for the residual mask effects in the power spectrum). Our estimators account for a number of systematic effects in both the power spectrum and bispectrum including radial integral constraints, stochasticity, and wide-angle effects. We additionally excise pairs of points with separations $\theta<0.05\degree$ from the estimators to mitigate fiber-collisions, using a pair-counting approach for the power spectrum \citep{Pinon:2024wzd} and a stochastic algorithm for the bispectrum \citep{desi1}. Finally, we compute Gaussian covariance matrices, which account for the mask, integral constraints, fiber collisions, systematics marginalization and Monte Carlo noise from the estimator \citep{Philcox:2024rqr,desi1}.

For the power spectrum, we include the monopole, quadrupole, and hexadecapole moments ($\ell=0,2,4$), restricting to $k\in[0.02,0.20]\hMpc$ with $\delta k = 0.01\hMpc$. The lower limit is chosen to minimize systematic contamination \citep{Rosado-Marin:2024xte,DESI:2024jis}, whilst the upper limit balances theoretical systematics and the shot-noise plateau. For the bispectrum, we restrict to the monopole moment ($\ell=0$) and utilize bins in the range $k\in[0.02,0.08]\hMpc$ following \citep{Ivanov:2021kcd}. We adopt a lower $k_{\rm max}$ for the bispectrum since we restrict to a tree-level theory model. Somewhat stronger constraints may be obtained using a higher-order model \citep[e.g.,][]{DAmico:2022osl,Philcox:2022frc,Bakx:2025pop} and by including the anisotropic bispectrum moments \citep{Ivanov:2023qzb}.

\subsection{External Data}\label{sec:data2}

\paragraph{BAO}
\noindent As in \paperone, we supplement our DR1 FS measurements with the official DESI DR2 BAO parameters \citep{DESI:2025zpo,DESI:2025zgx}. These constrain the evolution of cosmic distances and can be considered approximately independent from the full-shape statistics given that (a) they involve post-reconstructed statistics, due to which the BAO and FS datasets usually exhibit only weak correlations~\citep{Philcox:2020vvt}, and (b) the DR2 data contain many more galaxies than DR1, and its footprint is a factor of three larger. Explicit tests carried out in \paperone showed that the covariance between DR2 BAO and DR1 FS can safely be neglected. We include BAO data from each of the six chunks described above in addition to lower-redshift ELGs (omitted from DR1 due to systematic contamination \citep{DESI:2024jis}) and high-redshift Lyman-$\alpha$ tracers.

\vskip 6pt
\paragraph{CMB}
\noindent 
To break degeneracies intrinsic to late-time datasets, we include CMB information. Specifically, we utilize the high-$\ell$ \textsc{plik} TT, TE, and EE spectra, low-$\ell$ \textsc{SimAll} EE  and low-$\ell$ \textsc{Commander} TT likelihoods from the official \textsc{plik} 2018 release~\cite{Planck:2018nkj}. In addition, we use measurements of the lensing potential auto-spectrum $C_\ell^{\phi\phi}$ from \textit{Planck} PR4 and ACT DR6 \cite{Carron:2022eyg,ACT:2023kun,ACT:2023dou}.
We will refer these CMB datasets simply as ``CMB''. Where relevant, we also report the constraints using the latest \texttt{HiLLiPoP}+\texttt{LoLLiPoP} likelihoods~\cite{Tristram:2020wbi,Tristram:2023haj} based on the \textit{Planck} PR4 maps, which use slightly more data and have larger sky fractions, than \textsc{plik}. We combine the PR4 likelihoods with the low-$\ell$ \textsc{Commander} TT likelihood and the lensing potential measurements from \textit{Planck} PR4 and ACT DR6 data, as before.
We will denote these CMB measurements as ``$\cmbpr$''.

\vskip 6pt
\paragraph{SNe}
\noindent 
Low-redshift supernovae (SNe) provide a powerful geometric probe of the late-time Universe through their inferred distance-redshift relation. Following DESI \citep{DESI:2024mwx}, we consider three collations of supernovae: Pantheon\texttt{+} \citep{Brout:2022vxf}, Union3 \citep{Rubin:2023ovl}, and DES-SN5YR \citep{DES:2024jxu}. Each contains $\mathcal{O}(1000)$ Type Ia events with either photometric (DES-SN5YR) or spectroscopic (Pantheon\texttt{+} and Union3) confirmation, spanning a redshift range $0.01<z<2.26$. We caution that the three samples are not independent. We will label these supernova datasets as Pantheon+, Union3 and DESY5, respectively.
At the final preparation stage of this work, the updated DES-SN5YR data release became available~\citep{DES:2025sig}. In this work, we employ the pre-updated distance measurements from DES-Y5~\citep{DES:2024jxu}.

\subsection{Theory Model}\label{sec:data3}
\noindent In this work, we investigate several extensions to the standard cosmological model. These comprise: spatial curvature, parametrized by $\Omega_k$; non-minimal neutrino masses, described by $M_\nu\equiv \sum m_\nu$; dynamical dark energy, with equation-of-state $w(a) = w_0 + w_a(1-a)$. 
When the total neutrino mass is varied we assume three degenerate mass eigenstates~\footnote{This model provides an accurate approximation for the observable effects of both normal and inverted neutrino mass hierarchy and recovers the correct value of $M_\nu$ without any bias~\cite{Lesgourgues:2006nd}.} and impose a physical prior $M_\nu\geq 0$; when it is not varied, $M_\nu$ is fixed to the minimal value of $0.06\eV$, modeled as a single massive and two massless states.
The linear-order evolution of these is computed using the \textsc{class} code \citep{Lesgourgues:2011re}, which is sufficient to describe the CMB, BAO and SNe observables. This depends on six baseline parameters: $H_0,\omega_c\equiv \Omega_ch^2, \omega_b \equiv \Omega_bh^2, A_s, n_s,\tau$ (describing the expansion rate, physical dark matter and baryon density, primordial amplitude and slope and reionization optical depth respectively), as well as those describing the above extensions. To reduce parameter degeneracies, we fix $\tau$ to the CMB best-fit and impose a BBN prior on $\omega_b$ and a wide prior on $n_s$ when analyzing data combinations without the CMB, following the DESI official analysis~\cite{DESI:2024hhd}. In the BAO-only analyses, we additionally fix $n_s$ and $A_s$, since these parameters cannot be measured from the growth history alone.

To model the galaxy clustering dataset, we utilize a theory model based on the EFT, as described in \paperone and a number of previous works \citep[e.g.,][]{Philcox:2021kcw,Ivanov:2021kcd,Ivanov:2019pdj,Philcox:2021kcw,Ivanov:2021kcd,Chudaykin:2024wlw} (see \citep{Ivanov:2019pdj,DAmico:2019fhj,Chen:2021wdi,Vlah:2015zda,Pajer:2013jj,Carrasco:2012cv,Mercolli:2013bsa,Porto:2013qua,Vlah:2015sea,Senatore:2014via,Senatore:2014vja,Senatore:2014eva,Angulo:2015eqa,Senatore:2017hyk,Baumann:2010tm,Assassi:2014fva,Assassi:2015jqa,Chen:2020zjt,Blas:2015qsi,Blas:2016sfa,McDonald:2006mx,McDonald:2009dh,Ivanov:2022mrd} for former works), as detailed below. This is implemented in the \textsc{class-pt} code \citep{classpt}.\footnote{Several alternative codes now exist, including \textsc{pybird}, \textsc{velocileptors}, \textsc{class one-loop}, \textsc{folps} and \textsc{pbj} \citep{DAmico:2020kxu,Chen:2020fxs,Linde:2024uzr,Noriega:2022nhf,Moretti:2023drg}.} For the power spectrum, our model is similar to that used in the official DESI DR1 analysis \citep{DESI:2024jis,Maus:2024dzi}, and is computed at one-loop order in perturbations, including all relevant effects such as bias, resummation of long-wavelength displacements, corrections for short-wavelength physics, and departures from Poisson statistics. For the bispectrum, we adopt a tree-level model, restricting our attention to large-scales.

Tab.~\ref{tab:priors} summarizes the EFT parameters and priors employed in this work.
In full, our theory model contains fourteen `nuisance' parameters for each redshift chunk, three of which enter only in the bispectrum model. Via linearity, all except three can be marginalized over analytically, implying that our full analysis contains $6\times 3 = 18$ parameters in addition to those describing the cosmological model. 
Compared to the official DESI analyses~\citep{DESI:2024jis,DESI:2024hhd}, our EFT framework offers greater flexibility: it allows the cubic bias parameter $b_{\Gamma_3}$, the stochastic counterterm $a_0$ and the next-to-leading-order counterterm parameter $\tilde c$ to vary. These parameters have been detected in simulations~\cite{Ivanov:2024hgq,Chudaykin:2024wlw,Ivanov:2024xgb,Ivanov:2024dgv,Ivanov:2025qie}, and correspond to
important physical effects, for details see \paperone.
As described in \citep{Maus:2024dzi,desi1}, we can limit projection (`prior volume') effects by suitable model reparameterization; this is described in \citep{desi1} and requires us to sample $b_1\sigma_8$, $b_2\sigma_8^2$, $b_{\mathcal{G}_2}\sigma_8^2$ instead of the physical parameters $b_1,b_2,b_{\mathcal{G}_2}$. We utilize uninformative flat priors on all cosmological parameters\footnote{When analyzing dynamical dark energy, we enforce the physical constraint $w_0+w_a\leq 0$ following \citep{DESI:2024mwx}.} and $b_1\sigma_8$ assuming broad Gaussian priors for the remaining nuisance parameters. Our Gaussian priors build upon the analysis pipeline developed in \paperone, with several improvements introduced in this work. First, the next-to-leading-order counterterm parameter, $\tilde c$, (associated with fingers-of-God effects) and the cubic tidal bias, $b_{\Gamma_3}$, are rescaled by $\sigma^2_8(z)$ and $\sigma^4_8(z)$, respectively. This provides a more natural parametrization as the resulting parameter combinations appear directly in the theoretical model. Second, motivated by \citep{Tsedrik:2025hmj}, we rescale the nuisance parameters that multiply the linear power spectrum by the Alcock–Paczynski (AP)~\cite{Alcock:1979mp} amplitude. This parametrization accounts for the impact of the fiducial cosmology used to convert redshifts into distances in the data, which also modifies the overall amplitude of the signal. The resulting AP-rescaled combinations are more directly linked to the observed clustering signal, e.g., for the linear power spectrum, $P_\ell\propto A_{\rm AP}\sigma_8^2$, making them a more appropriate basis for defining EFT priors~\cite{Tsedrik:2025hmj}.~\footnote{We do not rescale $b_1$, $b_2$, and $b_{\mathcal{G}_2}$ with the AP amplitude because these bias parameters enter the theoretical model in different combinations with $A_{\rm AP}$. All other nuisance parameters appear in unique combinations with the AP amplitude, motivating our AP-rescaled priors.} We find that this reparameterization reduces marginalization shifts in parameter constraints caused by prior-volume projection effects in the $\wa$ model. The consistency tests of the updated analysis pipeline are presented in App.~\ref{app:marg}.

\begin{table}[t] 
    \centering
    \begin{tabular}{|lllll|}
    \hline
    Type & Parameter & Default & Prior & Units \\  
    \hline
    \textbf{nuisance} 
    & $b_1 \sigma_8(z)$ &  & $\mathcal{U}[0, 3]$ &---\\  (sampled) 
    & $b_2 \sigma_8^2(z)$ &  & $\mathcal{N}[0, 5^2]$ &---\\    
     & $b_{\mathcal{G}_2} \sigma_8^2(z)$ &  & $\mathcal{N}[0, 5^2]$ &---\\ 
    \hline
    \textbf{nuisance} & $b_{\Gamma_3}\,A_{\rm AP}\,A_{\rm amp}^2$ &  & $\mathcal{N}\left(\frac{23}{42}(b_1-1),1^2\right)$ &---\\
    (analytically & $c_0\,A_{\rm AP}\,A_{\rm amp}$ &  & $\mathcal{N}(0,30^2)$ & $[\Mpc/h]^2$\\
    marginalized) & $c_2\,A_{\rm AP}\,A_{\rm amp}$ &  & $\mathcal{N}(30,30^2)$ & $[\Mpc/h]^2$\\
    & $c_4\,A_{\rm AP}\,A_{\rm amp}$ &  & $\mathcal{N}(0,30^2)$ & $[\Mpc/h]^2$\\
    & $\tilde{c}\,A_{\rm AP}\,A_{\rm amp}$ &  & $\mathcal{N}(400,400^2)$ & $[\Mpc/h]^4$\\
    & $c_1\,A_{\rm AP}\,A_{\rm amp}$ &  & $ \mathcal{N}(0,5^2)$ & $[\Mpc/h]^2$\\
    & $P_{\rm shot}$ &  & $\mathcal{N}(0,1^2)$ &---\\
    & $a_{0}$ &  & $\mathcal{N}(0,1^2)$ &---\\
    & $a_{2}$ &  & $\mathcal{N}(0,1^2)$ &---\\
    & $B_{\rm shot}\,A_{\rm AP}\,A_{\rm amp}$ &  & $\mathcal{N}(0,1^2)$ &---\\
    & $A_{\rm shot}$ &  & $\mathcal{N}(0,1^2)$ &---\\
   \hline 
    \end{tabular}
    \caption{\textbf{Model Parameters}: 
    Parameters and priors used in the galaxy clustering analyses. Here, $\mathcal{U}[a,b]$ refers to a uniform prior between $a$ and $b$, whilst $\mathcal{N}(\mu,\sigma^2)$ denotes a Gaussian distribution with mean $\mu$ and variance $\sigma^2$. Bias parameters $b_1\sigma_8$, $b_2\sigma_8^2$, $b_{\mathcal{G}_2}\sigma_8^2$ are directly sampled in the MCMC chains, whilst the nuisance parameters that appear quadratically in the likelihood are marginalized over analytically, as discussed in the text. 
    We use the Alcock-Paczynski parameter, $A_{\mathrm{AP}}\equiv 
\left( \frac{H^{\mathrm{fid}}_0}{H_0} \right)^3 
\frac{H(z)}{H^{\mathrm{fid}}(z)}
\left( \frac{D^{\mathrm{fid}}_A(z)}{D_A(z)} \right)^2$, where the super-script ``fid'' refers to quantities evaluated in the fiducial cosmology assumed when converting redshifts to distances in the DESI data~\citep{desi1}. The $A_{\rm AP}$ factor is absorbed into the definition of the stochastic parameters.
    Similarly, we define $A_{\rm amp}\equiv\sigma_8^2(z)/\sigma_{8,{\rm ref}}^2(z)$, where $\sigma_{8,{\rm ref}}^2(z)$ is the late-time fluctuation amplitude at the \textit{Planck} 2018 best-fit cosmology~\cite{Planck:2018nkj}. 
    }
    \label{tab:priors}
\end{table}

\resub{
We model the effects of neutrinos on galaxy clustering
using the Einstein-de-Sitter (EdS) approximation, formulating the EFT expansion in terms of the
linear matter power spectrum for the ``cold dark matter+baryons'' (CDM+b) fluid.
It is known, however, that massive neutrinos induce a scale-dependent bias on large scales.
This results in a step-like feature in the linear bias of size $\sim f_\nu=\Omega_\nu/\Omega_m$, which partially compensates the neutrino-induced suppression in the matter power spectrum~\cite{LoVerde:2014pxa,LoVerde:2014rxa,Munoz:2018ajr}.
In this work, we do not incorporate such a scale-dependent bias, assuming the EFT parameters take $k$-independent values. 
This is a good approximation within the precision of current data since we adopt the bias expansion with respect to CDM+b, as opposed to all matter, which significantly suppress the scale-dependent linear bias feature~\cite{Villaescusa-Navarro:2013pva,Castorina:2013wga,Costanzi:2013bha,LoVerde:2013lta,LoVerde:2014pxa,Castorina:2015bma,LoVerde:2016ahu}. 
In particular, for $M_\nu=0.09\eV$, numerical calculations of the scale-dependent halo bias yields a step-like feature with amplitude $\Delta b_1/b_1\approx0.35\%$~\cite{Munoz:2018ajr}, which lies beyond the precision of current LSS measurements. Moreover, although the shape of this step is different from the $k^2$ scaling induced by small-scale feedback, including baryonic effects, its effect can be partially absorbed by EFT counterterms on large scales~\cite{Munoz:2018ajr}. 
Massive neutrinos also induce a scale dependence in the logarithmic growth factor, whose impact on clustering statistics is likewise suppressed when adopting the CDM+b prescription~\cite{Villaescusa-Navarro:2017mfx}. Although it will be important to model corrections from the scale-dependent galaxy bias and growth rate accurately in the future, these effects do not affect the conclusions of this work.
}

We perform Markov Chain Monte Carlo (MCMC) analyses to sample from the posterior distributions using the Metropolis-Hastings algorithm as implemented in the \textsc{Montepython} code~\cite{Audren:2012wb,Brinckmann:2018cvx}. Marginalized constraints are computed and visualized using the public \textsc{getdist} code \citep{Lewis:2019xzd}.\footnote{\href{https://getdist.readthedocs.io/en/latest/}{https://getdist.readthedocs.io/en/latest/}} We adopt a Gelman-Rubin~\cite{Gelman:1992zz} convergence criterion $R-1<0.02$ for all analyses in this work. To determine the best-fit points and the corresponding $\chi^2_{\rm MAP}$, we use the simulated-annealing optimizer \textsc{Procoli} package~\cite{Karwal:2024qpt}.\footnote{\href{https://github.com/tkarwal/procoli}{https://github.com/tkarwal/procoli}}
We extract the best-fit parameters by minimizing the likelihood with the 
Jeffreys prior (i.e. dropping the determinant term in the marginalized likelihood),
which is  
equivalent to minimizing the full 
un-marginalized 
likelihood.

\section{Results}\label{sec:results}
\noindent
We now present the constraints on non-zero spatial curvature $o\Lambda$CDM, the dynamical dark energy model $w_0w_a$CDM and the minimal cosmological model $\Lambda$CDM supplemented with free
neutrino masses, dubbed $\Lambda$CDM+$M_\nu$. As detailed before, $\Lambda$CDM+$M_\nu$ is the minimal extension motivated by the fact that the neutrino masses are the known unknown parameters of our Universe. The non-zero spatial curvature and dynamical dark energy are the `classical' extensions for which the cosmological datasets have shown some preference in the past, e.g.~\cite{Aghanim:2018eyx}. Most recently, DESI reported some preference for the dynamical dark energy and the negative curvature~\cite{DESI:2024mwx,DESI:2025zgx}. Given these preferences and their impact on neutrino mass inference, we explore the constraints on neutrino masses in the $\wa$ and $\OmK$ backgrounds. 

\resub{Before presenting the results, it is useful to assess the sensitivity of the full-shape observables to parameters describing non-minimal cosmological scenarios. Since the effects of massive neutrinos on clustering statistics is well understood (see e.g.,~\cite{Lesgourgues:2006nd}), we focus here on the $\OmK$ and $\wa$ models. As a first step, we compute the best-fit $\ld$ model from the $\cmb+P_\ell+B_0+\bao$ data by fitting the six cosmological parameters together with six copies of nuisance parameters, following the procedure described in Sec.~\ref{sec:data3}. The EFT parameters that enter quadratically in the likelihood are recovered {\it a posteriori} from the MCMC chains. This best-fit model is then used to fix the standard cosmological parameters $\{\omega_{cdm}, \omega_{b}, \theta_*, \ln(10^{10}A_s), n_s,\tau\}$ as well as the nuisance parameters $\{b_1,b_2,b_{\mathcal{G}_2},b_{\Gamma_3},c_0, c_2,c_4,b_4,P_{\rm shot},a_0,a_2\}$ corresponding to the LRG3 sample. Importantly, we fix the CMB acoustic scale, $100\theta_*=1.0421$, which approximately incorporates CMB information in non-standard cosmological scenarios.}
\resub{In the $\OmK$ model, we explicitly vary $\Omk$, while in the $\wa$ scenario we vary $w_0$ and fix $w_a=0$ for simplicity.}
\resubb{Note that only the present-day physical densities of cold dark matter and baryons are held fixed, while $\Omega_m$ and $H_0$ are allowed to vary in each scenario to reproduce the measured value of $100\theta_*$.}

\begin{figure}[!t]
	\centering
    \includegraphics[width=0.48\textwidth]{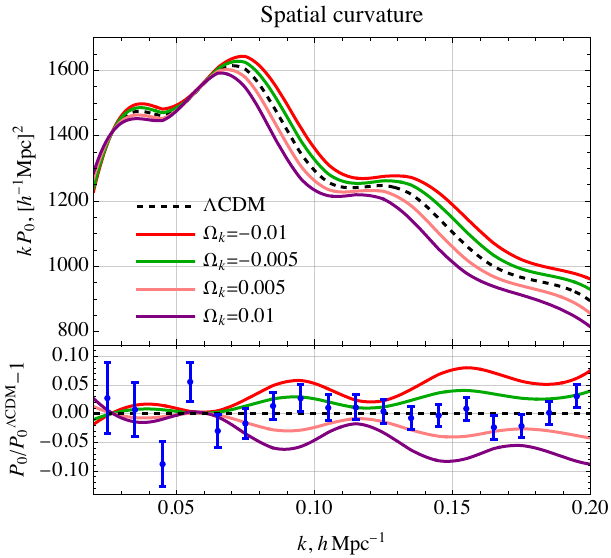}
    \includegraphics[width=0.48\textwidth]{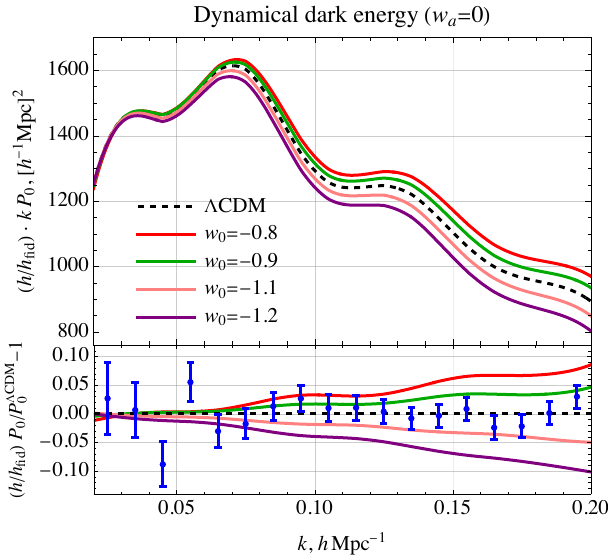}
	\caption{\resub{Theoretical predictions for the galaxy power spectrum monopole at $z_{\rm eff}=0.922$ for the $\OmK$ model (left panel) and $\wa$ scenario (right panel), with the latter evaluated with $w_a=0$. The galaxy power spectrum is modeled using the one-loop EFT framework fixing the CMB acoustic scale $\theta_*$ and all other cosmological and bias parameters; see the text for details. 
    The lower panel displays the residuals relative to the $\ld$ model, together with the DESI DR1 measurements from the LRG3 galaxy sample (blue points with $1\sigma$ error-bars).
    To illustrate the impact of $w_0$ on the shape of the galaxy power spectrum, we rescale the theoretical prediction by $(h/h_{\rm fid})$, where $h_{\rm fid}$ corresponds to the Hubble constant in the $\ld$ model. 
    }}
    \label{fig:sens}
\end{figure}

\resub{Fig.~\ref{fig:sens} shows the impact of $\Omk$ and $w_0$ parameters on the galaxy power spectrum monopole evaluated at $z_{\rm eff}=0.922$. Lower panel displays the residuals with respect to the best-fit $\ld$ model, together with the DESI DR1 measurements from the LRG3 galaxy sample. Although must carefully account for the degeneracies between cosmological and bias parameters, it is clear that changing both $\Omk$ and $w_0$ leads to marked (and distinguishable) changes in the observed power spectra, particularly on small scales.}

\subsection{Spatial Curvature}\label{sec:OmK}
\begin{table*}[!t]
    \centering
    \begin{tabular}{lccccc}
    \toprule
    Dataset 
    & $10^3\Omk$ 
    & $\Omega_m$ 
    & $H_0$ 
    & $\sigma_8$  
    & $\enspace p(\Omk>0)\enspace $
    \\
    \midrule
    $\bao$
    & $\enspace 21.0_{-40.3}^{+40.2}\enspace$ 
& $\enspace 0.294_{-0.012}^{+0.012}\enspace$ 
& $\enspace 67.77_{-1.68}^{+1.68}\enspace$ 
& $\enspace 0.765_{-0.086}^{+0.075}\enspace$ 
& $0.5\sigma$
    \\
    $\bao+P_\ell$ 
    & $1.7_{-24.8}^{+22.5}$ 
& $0.2927_{-0.0074}^{+0.0074}$ 
& $68.64_{-1.08}^{+1.09}$ 
& $0.830_{-0.035}^{+0.036}$ 
& $<0.1\sigma$
    \\
    $\bao+P_\ell+B_0$ 
    & $-8.2_{-21.7}^{+21.6}$ 
& $0.2974_{-0.0069}^{+0.0070}$ 
& $69.14_{-1.00}^{+1.01}$ 
& $0.812_{-0.031}^{+0.032}$ 
& $-0.4\sigma$
    \\\hline
    $\cmb$ 
    & $-10.3_{-5.7}^{+7.0}$ 
& $0.352_{-0.025}^{+0.021}$ 
& $63.59_{-2.18}^{+2.21}$ 
& $0.796_{-0.010}^{+0.012}$ 
& $-1.5\sigma$
    \\
    $\cmb+\bao$ 
    & $2.4_{-1.2}^{+1.2}$ 
& $0.3031_{-0.0038}^{+0.0039}$ 
& $68.64_{-0.32}^{+0.32}$ 
& $0.8157_{-0.0060}^{+0.0060}$ 
& $1.9\sigma$
    \\
    $\cmb+P_\ell$ 
    & $4.3_{-1.9}^{+1.9}$ 
& $0.2941_{-0.0068}^{+0.0068}$ 
& $69.69_{-0.77}^{+0.77}$ 
& $0.8183_{-0.0064}^{+0.0064}$ 
& $2.3\sigma$
    \\
    $\cmb+P_\ell+B_0$ 
    & $4.3_{-1.9}^{+1.9}$ 
& $0.2953_{-0.0067}^{+0.0068}$ 
& $69.60_{-0.76}^{+0.76}$ 
& $0.8177_{-0.0062}^{+0.0063}$ 
& $2.4\sigma$
    \\
    $\cmb+P_\ell+B_0+\bao$ 
    & $2.7_{-1.2}^{+1.2}$ 
& $0.3007_{-0.0036}^{+0.0036}$ 
& $68.87_{-0.31}^{+0.31}$ 
& $0.8148_{-0.0058}^{+0.0058}$ 
& $2.3\sigma$
    \\
    $\cmbpr+P_\ell+B_0+\bao$ 
    & $2.2_{-1.1}^{+1.1}$ 
& $0.2996_{-0.0034}^{+0.0034}$ 
& $68.77_{-0.30}^{+0.30}$ 
& $0.8137_{-0.0051}^{+0.0052}$ 
& $1.9\sigma$
    \\
    \bottomrule
    \end{tabular}
    \caption{\textbf{Spatial curvature}: Mean and 68\% confidence intervals on cosmological parameters in the $\OmK$ analyses. The addition of the full-shape power spectrum ($P_\ell$) and bispectrum ($B_0$) to the BAO dataset leads to $\sim 2\times$ sharper constraints on $\Omk$ in DESI-only analyses, but does not lead to significant information gain when the CMB results are included. \resub{The last column presents the deviation from $\Omk=0$, determined by the fraction of the posterior
enclosing $\Omk>0$ (for negative curvature) or $\Omk<0$ (for positive curvature) expressed in terms of of a $\sigma$-interval.}
    }
    \label{tab:OmK}
\end{table*}
\begin{figure}[!t]
	\centering
    \includegraphics[width=0.44\textwidth]{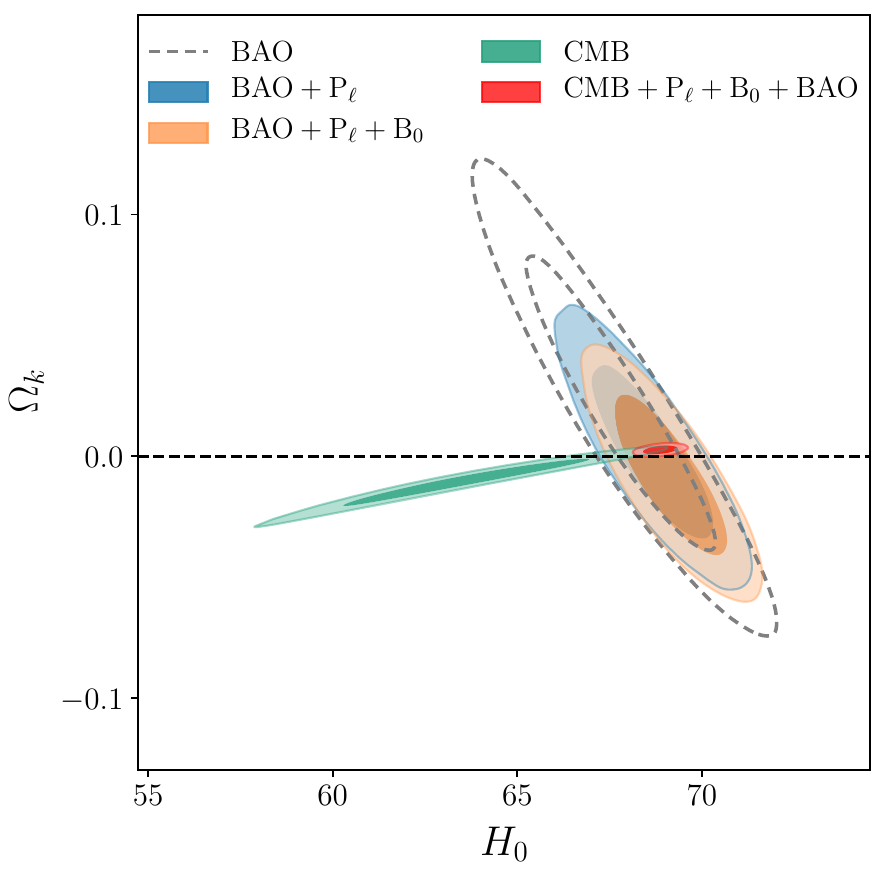}
    \includegraphics[width=0.44\textwidth]{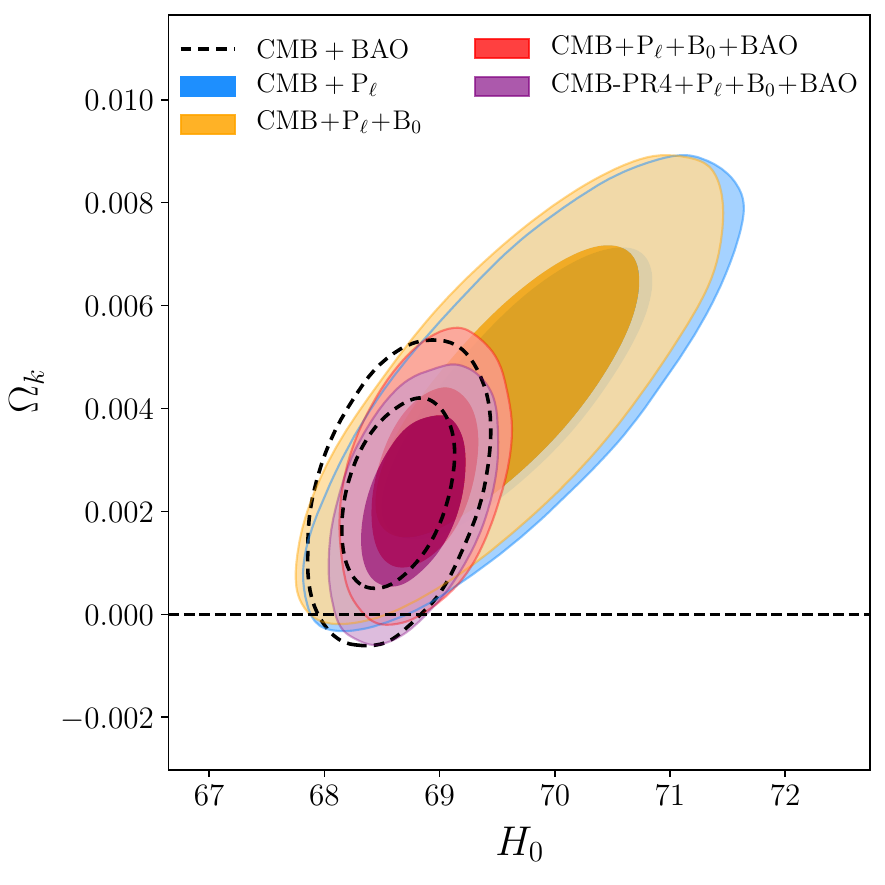}
	\caption{\textbf{Spatial curvature}: Two-dimensional posterior distributions in the $\Omk-H_0$ plane in the $\OmK$ analyses. {\it Left panel:} The inclusion of the DESI data breaks the geometric degeneracy present in the CMB data and brings $\Omk$ into better agreement with a flat universe. {\it Right panel:} The combined analyses of the CMB and DESI data show a mild preference for $\Omk>0$, with a significance ranging from $1.9\sigma$ to $2.4\sigma$, depending on the dataset combination. Using the newer \texttt{HiLLiPoP}+\texttt{LoLLiPoP} likelihoods~\cite{Tristram:2020wbi,Tristram:2023haj}, based on the Planck PR4 maps, alleviates this deviation, yielding $\Omk$ values consistent with zero within 95\% CL.
    }
    \label{fig:OmK}
\end{figure}

\noindent

We start by presenting the cosmological results with a free spatial curvature.
Tab.~\ref{tab:OmK} summarizes the parameter constraints from DESI alone as well as in combination with CMB, whilst Fig.~\ref{fig:OmK} shows the two-dimensional marginalized posterior distributions in the $\Omk-H_0$ plane for the selected $\OmK$ analyses. The left panel demonstrates the geometric degeneracy present in the CMB data, which is broken by the inclusion of the DESI BAO data, pulling the spatial curvature into closer agreement with a flat universe.
The right panel of Fig.~\ref{fig:OmK} illustrates the impact of the DESI DR1 power spectrum and bispectrum measurements. We find that the results of the combined CMB+BAO analyses are consistent with those that include FS information. 

Let us consider these results in more detail. The DESI BAO data provides a tomographic measurement of the expansion rate, which allows one to determine the $\Omk$ parameter without any external distance calibration. From the DESI DR2 BAO data alone, we obtain $10^3\,\Omk=21.0_{-40.3}^{+40.2}$, in excellent agreement with the official DESI analysis~\cite{DESI:2025zgx}. From the shape of the galaxy power spectrum, we can obtain an independent constraint on the physical dark matter density $\omega_{cdm}$, which sets the sound horizon at the drag epoch.\footnote{The sound horizon at the baryon drag epoch depends on both $\omega_b$ and $\omega_{cdm}$. Since the DESI data does not have sufficient statistical power to constrain the baryon density $\omega_b$, we adopt the BBN prior to constrain this parameter.} 
The FS measurements thus act as an external calibration of the sound horizon, breaking the $H_0-r_d$ degeneracy and constraining parameters that govern the expansion history. We find that incorporating the DESI DR1 pre-reconstructed power spectrum reduces the $\Omk$ uncertainty by a factor of two, with a further $10\%$ improvement found by adding the DESI DR1 bispectrum, due to parameter degeneracy breaking.
All of the DESI-only results are fully consistent with a flat Universe. 

The CMB provides an independent and precise measurement of spatial curvature. Using the CMB data alone, we find $10^3\,\Omk = -10.3_{-5.7}^{+7.0}$, which is roughly three times tighter than that from the $\bao+P_\ell+B_0$ analysis (see Fig.\,\ref{fig:OmK}). 
\resub{Notably, the CMB measurements imply a mild preference for positive curvature ($\Omk<0$) at the $1.5\sigma$ level.
This preference can be traced back to the so-called ``curvature tension'' in the Planck PR3 data~\cite{Planck:2018nkj}.
In our analysis, the evidence for positive curvature is somewhat reduced due to the inclusion of the latest CMB lensing likelihoods~\cite{Planck:2018nkj,Tristram:2023haj}.} 
The CMB constraint is robust to the choice of Planck likelihood: adopting the newer \texttt{HiLLiPoP}+\texttt{LoLLiPoP} likelihoods~\cite{Tristram:2020wbi,Tristram:2023haj}, based on the \textit{Planck} PR4 maps, yields $10^3\,\Omk = -8.7_{-5.6}^{+6.2}$, fully consistent with the Planck PR3 result. 
\resub{This preference for $\Omk<0$ in the CMB data is not expected to affect our conclusions because the addition of the BAO and FS data breaks parameter degeneracies and favours $\Omk>0$, as discussed below.}

Finally, we combine the CMB and DESI measurements. As shown in the left panel of Fig.~\ref{fig:OmK}, the degeneracy directions in the CMB and large-scale structure data are nearly orthogonal, suggesting a large information gain from combining these datasets. 
In the $\cmb+\bao$ analysis, the $\Omk$ errorbar shrinks by a factor of 5.5 compared to the CMB-only result. When the CMB data are combined with the DESI DR1 FS measurements, the improvement is somewhat smaller. Specifically, in the $\cmb+P_\ell$ analysis, the $\Omk$ uncertainty is reduced by a factor of 3.5 relative to the CMB-only case, and including the bispectrum does not significantly change the parameter constraints. Interestingly, adding the FS information slightly increases the preference for $\Omk>0$ \resub{(see Tab.~\ref{tab:OmK})}~\citep{DESI:2024mwx,DESI:2025zgx}.
The deviation from $\Omk=0$ corresponds to $1.9\sigma$, $2.4\sigma$, $2.4\sigma$, and $2.3\sigma$ for the $\cmb+\bao$, $\cmb+P_\ell$, $\cmb+P_\ell+B_0$, and $\cmb+P_\ell+B_0+\bao$ analyses respectively.\footnote{Hereafter, we evaluate the significance of parameter deviations using the actual posterior distribution, without assuming Gaussianity.}
However, the evidence for negative curvature in the joint analysis decreases to below $2\sigma$ when using the newer \texttt{HiLLiPoP}+\texttt{LoLLiPoP} CMB likelihoods~\cite{Tristram:2020wbi,Tristram:2023haj}.
We conclude that the deviation from $\Omk=0$ is not significant and fully consistent with being a statistical fluctuation. 

\resub{It is useful to compare our measurements of the weak-lensing parameter, $S_8\equiv\sigma_8\sqrt{\Omega_m/0.3}$, with those from the DES and KiDS datasets. Our analyses involving DESI measurements predict moderately larger values of $S_8$ compared to the joint DES Y3+KiDS-1000 analysis~\cite{Kilo-DegreeSurvey:2023gfr}, but remain consistent within $1.6\sigma$. When compared to the more recent KiDS-Legacy constraint~\cite{Wright:2025xka}, our measurements are consistent within $1\sigma$.
}

\subsection{Dynamical Dark Energy}\label{sec:w0wa}
\begin{figure}[!t]
	\centering
	\includegraphics[width=0.45\textwidth]{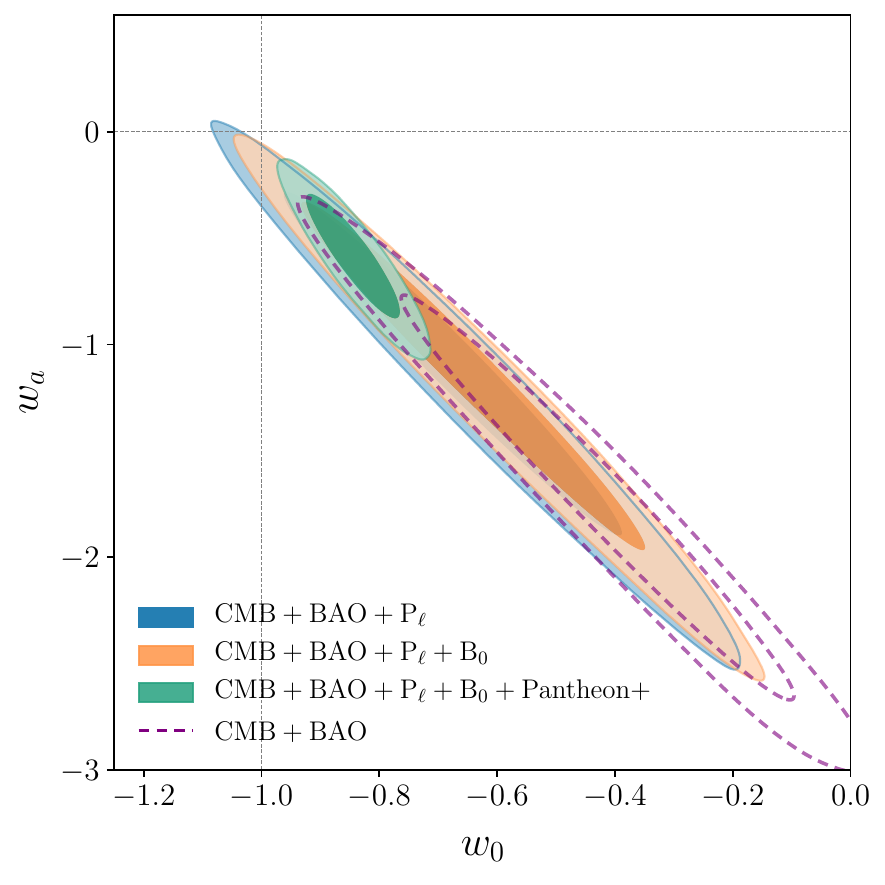}
    \includegraphics[width=0.45\textwidth]{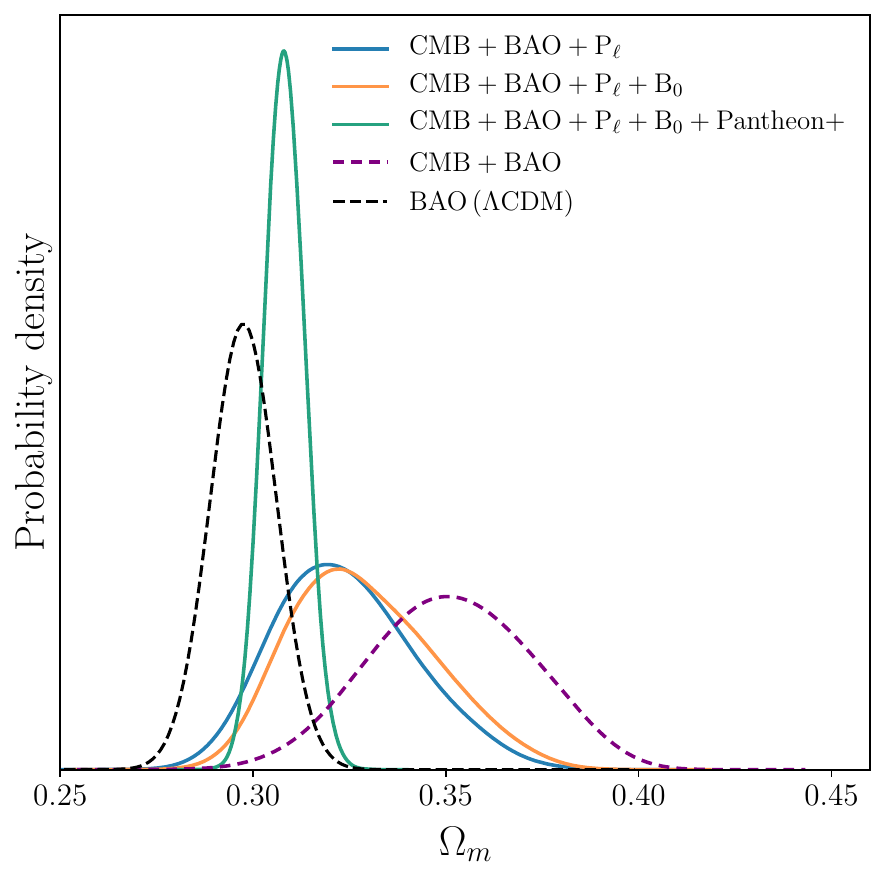}
	\caption{\textbf{Dynamical Dark Energy}: Two-dimensional posterior distributions in the $w_0-w_a$ plane ({\it left panel}) and one-dimensional marginalized posteriors for $\Omega_m$ ({\it right panel}), when fixing the background model to $\wa$. We also show the $\Omega_m$ posterior from the official DESI analysis of the BAO DR2 data in the $\ld$ model (black dashed line)~\cite{DESI:2025zgx}. Corresponding results including various SNe datasets are shown in Fig.\,\ref{fig:w0wa_sn}.
    {\it Left panel:} The significance of the tension with $\ld$ ($w_0=-1$, $w_a=0$) estimated from the $\Delta\chi^2_{\rm MAP}$ values is $2.9\sigma$, $2.8\sigma$ and $2.9\sigma$ for the $\rm \cmb+\bao$, $\rm \cmb+\bao+P_\ell+B_0$ and $\rm \cmb+\bao+P_\ell+B_0+Pantheon+$ analyses, respectively.
    {\it Right panel:} The addition of the DESI DR1 FS information shifts the $\Omega_m$ posterior towards lower values in a closer agreement with the DESI-only $\ld$ result.  
    }
    \label{fig:w0wa}
\end{figure}

\begin{figure*}[!t]
\includegraphics[width=0.325\textwidth]{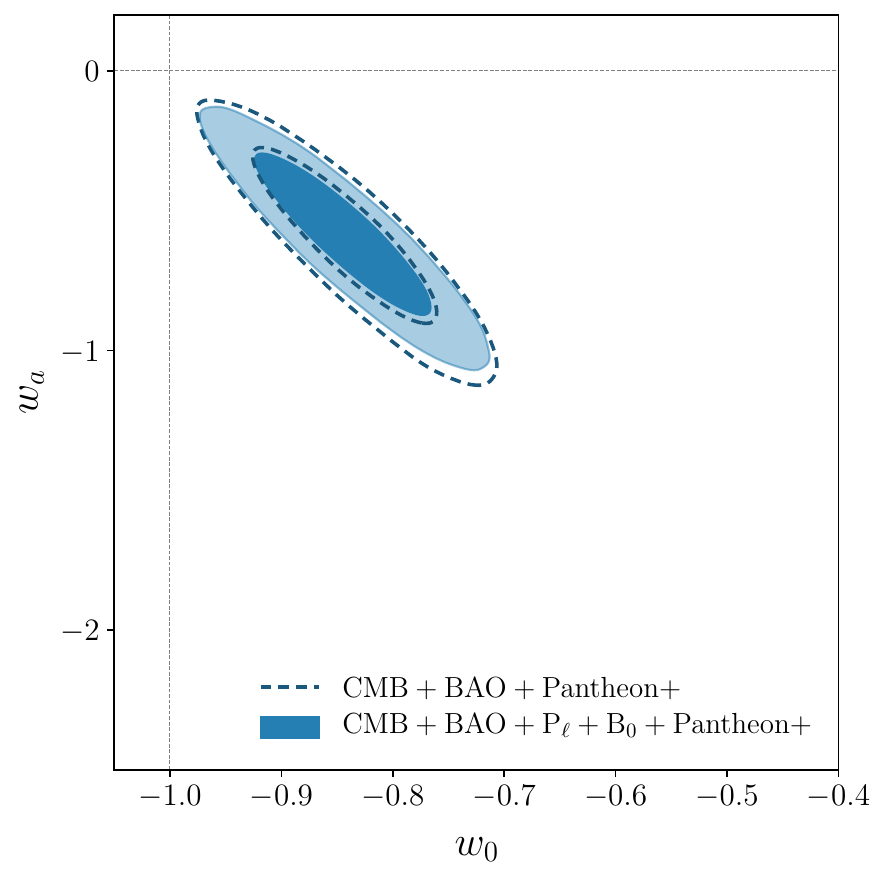}
\includegraphics[width=0.325\textwidth]{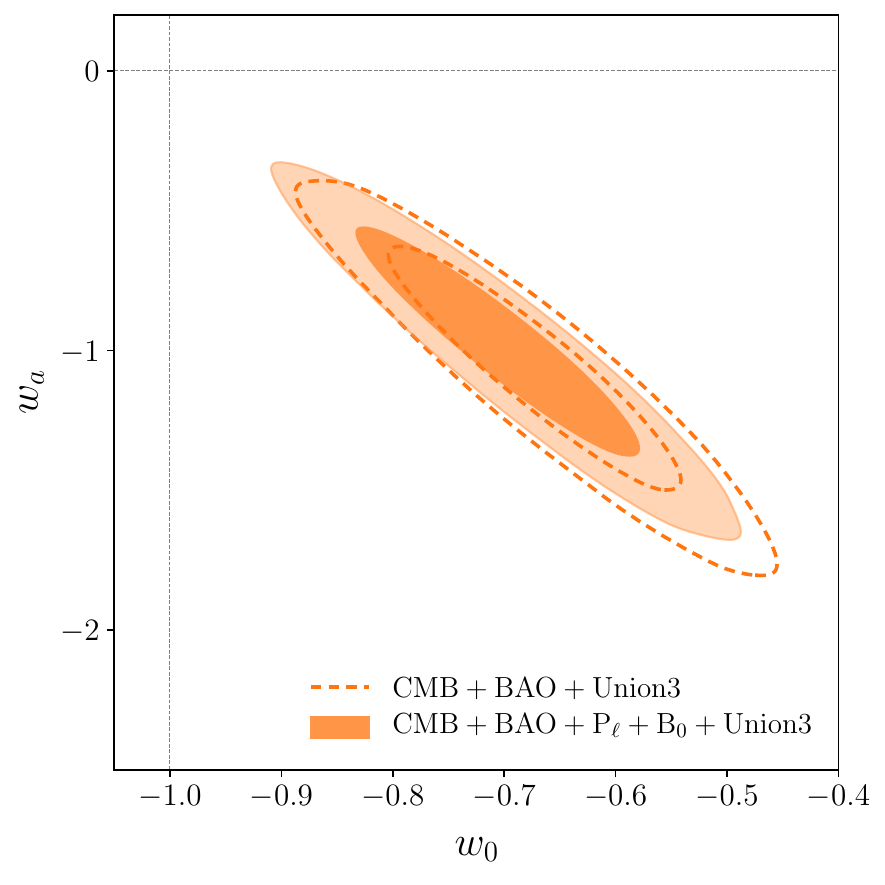}
\includegraphics[width=0.325\textwidth]{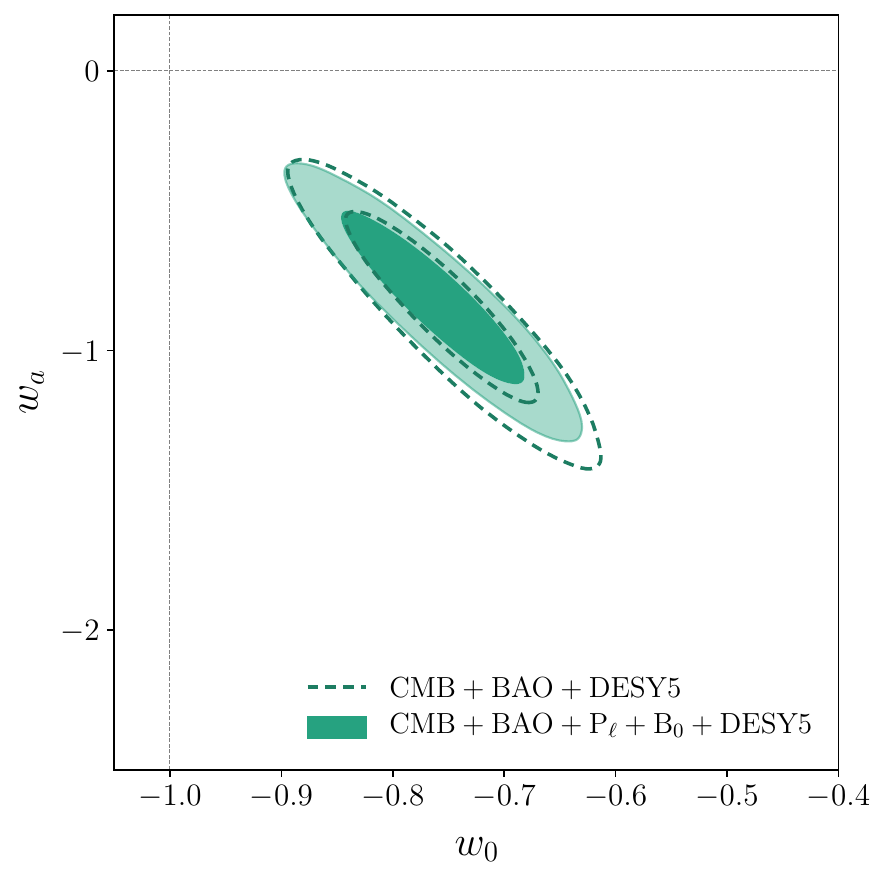}
\caption{
\textbf{Dynamical Dark Energy}: Two-dimensional marginalized constraints on $w_0$ and $w_a$ parameters in the $\wa$ analyses of the $\cmb+\bao+\sn$ (dashed) and $\rm \cmb+\bao+\sn+P_\ell+B_0$ (solid) data for three different SN Ia samples. The addition of the DESI DR1 power spectrum ($P_\ell$) and bispectrum monopole ($B_0$) data considerably improves constraints on the dark energy parameters: the Figure of Merit (FoM) for $w_0$ and $w_a$ increases by factors of $1.19$, $1.17$, and $1.19$ upon adding the DESI DR1 FS information for combinations involving PantheonPlus, Union3, and DESY5, respectively.
}
\label{fig:w0wa_sn} 
\end{figure*}

\noindent
In this section, we present cosmological constraints on the $\wa$ model. We adopt the CPL parameterization $w(a)=w_0+w_a(1-a)$ for the dark energy equation of state and compute perturbations using the parameterized post-Friedmann (PPF) approach~\cite{Fang:2008sn}. To ensure a matter-dominated epoch in the past, we impose the condition $w_0+w_a<0$. Tab.~\ref{tab:w0wa} lists the one-dimensional marginalized parameter constraints, whilst Fig.~\ref{fig:w0wa} shows the two-dimensional posterior distributions in the $w_0-w_a$ plane and the one-dimensional posteriors for $\Omega_m$. 
Fig.~\ref{fig:w0wa_sn} illustrates the impact of the DESI DR1 FS measurements on the $w_0$ and $w_a$ posteriors. 
\begin{table*}[!t]
    \centering
    \begin{tabular}{lcccccc}
    \toprule
    Dataset 
    & $w_0$ 
    & $w_a$ 
    & $\Omega_m$ 
    & $H_0$ 
    & $\sigma_8$  
    & $N_\sigma(w_0,w_a)$  
    \\
    \midrule
    $\cmb+\bao$
    & $\enspace -0.43_{-0.21}^{+0.22}\enspace$ 
& $\enspace -1.71_{-0.63}^{+0.61}\enspace$ 
& $\enspace 0.351_{-0.022}^{+0.022}\enspace$ 
& $\enspace 63.86_{-2.10}^{+1.70}\enspace$ 
& $\enspace 0.784_{-0.017}^{+0.015}\enspace$ 
& $3.0\sigma$
    \\
    $\cmb+\bao+{\rm Pantheon}+$  
    & $-0.840_{-0.055}^{+0.054}$ 
& $-0.60_{-0.19}^{+0.22}$ 
& $0.3108_{-0.0057}^{+0.0056}$ 
& $67.64_{-0.59}^{+0.59}$ 
& $0.8117_{-0.0085}^{+0.0085}$ 
& $2.6\sigma$
    \\
    $\cmb+\bao+{\rm Union3}$  
    & $-0.669_{-0.087}^{+0.087}$ 
& $-1.08_{-0.28}^{+0.31}$ 
& $0.3269_{-0.0086}^{+0.0086}$ 
& $66.05_{-0.83}^{+0.84}$ 
& $0.8009_{-0.0094}^{+0.0095}$ 
& $3.5\sigma$
    \\
    $\cmb+\bao+{\rm DESY5}$  
    & $-0.753_{-0.057}^{+0.057}$ 
& $-0.85_{-0.21}^{+0.24}$ 
& $0.3186_{-0.0056}^{+0.0056}$ 
& $66.87_{-0.56}^{+0.56}$ 
& $0.8073_{-0.0081}^{+0.0081}$ 
& $4.7\sigma$
    \\\hline
    $\cmb+\bao+P_\ell$ 
    & $-0.66_{-0.21}^{+0.16}$ 
& $-1.14_{-0.45}^{+0.61}$ 
& $0.324_{-0.021}^{+0.016}$ 
& $66.34_{-1.71}^{+1.96}$ 
& $0.802_{-0.014}^{+0.016}$ 
& $2.6\sigma$
    \\
    $\cmb+\bao+P_\ell+B_0$ 
    & $-0.63_{-0.22}^{+0.16}$ 
& $-1.19_{-0.45}^{+0.62}$ 
& $0.329_{-0.022}^{+0.016}$ 
& $65.85_{-1.71}^{+2.00}$ 
& $0.796_{-0.014}^{+0.016}$ 
& $2.4\sigma$
    \\
    $\cmb+\bao+P_\ell+B_0+{\rm Pantheon}+$  
    & $-0.842_{-0.052}^{+0.052}$ 
& $-0.59_{-0.19}^{+0.19}$ 
& $0.3081_{-0.0053}^{+0.0053}$ 
& $67.84_{-0.55}^{+0.55}$ 
& $0.8105_{-0.0076}^{+0.0077}$ 
& $2.8\sigma$
    \\
    $\cmb+\bao+P_\ell+B_0+{\rm Union3}$  
    & $-0.699_{-0.089}^{+0.081}$ 
& $-0.99_{-0.26}^{+0.29}$ 
& $0.3214_{-0.0087}^{+0.0079}$ 
& $66.50_{-0.82}^{+0.82}$ 
& $0.8012_{-0.0086}^{+0.0087}$ 
& $3.8\sigma$
    \\
    $\cmb+\bao+P_\ell+B_0+{\rm DESY5}$  
    & $-0.761_{-0.054}^{+0.054}$ 
& $-0.82_{-0.20}^{+0.20}$ 
& $0.3155_{-0.0054}^{+0.0054}$ 
& $67.09_{-0.54}^{+0.54}$ 
& $0.8055_{-0.0074}^{+0.0074}$ 
& $4.2\sigma$
    \\
    \bottomrule
    \end{tabular}
    \caption{\textbf{Dynamical Dark Energy}: Mean and 68\% confidence intervals on cosmological parameters for the $\wa$ analyses. The two-dimensional posteriors on $(w_0,w_a)$ are shown in Fig.\,\ref{fig:w0wa}. We find enhanced constraints when adding the DESI DR1 full-shape power spectrum ($P_\ell$) and bispectrum ($B_0$) datasets, both with and without SNe. \resub{In the last column, we report the deviation from the $\Lambda$CDM model $(w_0,w_a)=(-1,0)$. The significance is derived from the marginalized two-dimensional posterior in the $(w_0,w_a)$ plane by determining the highest-posterior-density contour that passes through the $\Lambda$CDM point and converting the enclosed posterior probability into an equivalent one-dimensional Gaussian $N\sigma$ confidence interval.
    }
    }
    \label{tab:w0wa}
\end{table*}

The cosmological constraints derived from the CMB and DESI DR2 BAO datasets have already been presented by the DESI collaboration~\cite{DESI:2025zgx}. Here, we focus on the cosmological implications of the DESI DR1 FS measurements.
We begin by adding the FS information to the $\cmb+\bao$ combination. The addition of the pre-reconstructed power spectrum ($P_\ell$) pulls the $w_0$ and $w_a$ posteriors towards the $\ld$ values ($w_0=-1$, $w_a=0$), making them more consistent with a cosmological constant. In particular, in the $\cmb+\bao$ analysis, the $w_a$ posterior deviates from zero at the $3.3\sigma$ level, but this deviation decreases to $2.4\sigma$ when the DESI DR1 power spectrum is included. 
The Figure of Merit (FoM)~\cite{Albrecht:2006um}\footnote{The Figure of Merit (FoM) is defined as an inverse area of the posterior, ${\rm FoM}=|{\rm det} \mathbf{C}|^{-1/2}$, where $\mathbf{C}$ is the projected covariance matrix in the $(w_0,w_a)$ parameter space.} in the $w_0-w_a$ parameter plane increases by more than $30\%$ upon adding the DESI DR1 FS information.
An important implication of the FS measurements is that they provide an independent constraint on $\Omega_m$, as illustrated in the right panel of Fig.~\ref{fig:w0wa}. 
We find that adding the DESI DR1 power spectrum tightens the $\Omega_m$ posterior by $15\%$ relative to the CMB+BAO result.
The central value of $\Omega_m$ shifts lower by $1.3\sigma$, bringing it into closer agreement with the DESI-only $\ld$ result, $\Omega_m=0.2975\pm0.0086$~\cite{DESI:2025zgx}. 
Interestingly, the inclusion of the bispectrum does not improve the precision of parameter measurements in the $\wa$ scenario.\footnote{This behavior can be attributed to the adopted EFT priors. When using the original priors from \paperone, adding the higher-order statistics improves cosmological constraints by $\lesssim10\%$, in full agreement with \citep{desi1}; see App.~\ref{app:old} for details.} 

We emphasize that these results present the first cosmological constraints obtained from the CMB and DESI BAO+FS datasets without SNe. Earlier analyses were significantly affected by parameter projection effects (see Appendix A of~\cite{DESI:2024hhd}); to mitigate this, the official DESI collaboration $\wa$ analyses included the BAO+FS dataset only in combination with SNe. In contrast, our analysis pipeline produces robust results even without supernova distance information, thanks to the reparameterization of the EFT parameters described in Sec.~\ref{sec:data3}. This is validated with parameter recovery tests presented in Appendix~\ref{app:marg}.

Next, we discuss the results including SNe information. The addition of the DESI DR1 power spectrum and bispectrum increases the precision of one-dimensional marginalized parameter constraints by approximately $10\%$ relative to the combined CMB, BAO and SNe analysis. The FoM for $w_0$ and $w_a$ increases by $\approx 20\%$ for all three supernova datasets when the FS data are added, as shown in Fig.~\ref{fig:w0wa_sn}. 
Our results are consistent with those of the official DESI DR1 analysis~\cite{DESI:2024hhd}, but yield tighter constraints due to the use of the newer DESI DR2 BAO measurements in our study.
\resub{The DESI analyses predict moderately larger values of $S_8$ compared to the joint DES Y3+KiDS-1000 analysis~\cite{Kilo-DegreeSurvey:2023gfr}, while remaining consistent within $2.1\sigma$. At the same time, they are consistent within $1\sigma$ with the most recent KiDS-Legacy constraint~\cite{Wright:2025xka}, which yields a considerably higher value of $S_8$. }

Tab.~\ref{tab:chi2} lists the significance of the preference for the $\wa$ model over $\ld$, based on the $\Delta\chi^2_{\rm MAP}$, evaluated at the maximum a posteriori (MAP) points.
\begin{table}
\centering
    \begin{tabular}{lccc}
    \toprule
    Datasets & $\Delta\chi^2_{\rm MAP}$ & Significance  \\
    \midrule
    $\cmb+\bao$ 
    & $-10.8$ 
    & $2.8\sigma$ \\
    $\rm \cmb+\bao+P_\ell+B_0$ 
    & $-10.9$ 
    & $2.9\sigma$ \\\hline
    $\rm \cmb+\bao+Pantheon+$ 
    & $-9.8$ 
    & $2.7\sigma$ \\
    $\rm \cmb+\bao+Union3$ 
    & $-16.4$ 
    & $3.6\sigma$ \\
    $\rm \cmb+\bao+DESY5$ 
    & $-20.4$ 
    & $4.1\sigma$ \\
    $\rm \cmb+\bao+P_\ell+B_0+Pantheon+$ 
    & $-11.4$ 
    & $2.9\sigma$ \\
    $\rm \cmb+\bao+P_\ell+B_0+Union3$ 
    & $-17.5$ 
    & $3.8\sigma$ \\
    $\rm \cmb+\bao+P_\ell+B_0+DESY5$ 
    & $-22.3$ 
    & $4.3\sigma$ \\
    \bottomrule
    \end{tabular}
\caption{\textbf{Dynamical Dark Energy}: Difference in the $\chi^2_{\rm MAP}$ value for the best-fit $\wa$ model relative to the best-fit $\ld$ model with ($w_0=-1$, $w_a=0$) for fits to different combinations of datasets (as indicated). The third column lists the corresponding (frequentist) significance levels in terms of a $\sigma$-interval given the two additional free parameters.
\label{tab:chi2}
}
\end{table}
The addition of FS information \resub{slightly increases} the evidence for the $\wa$ model in the $\cmb+\bao$ analysis from $2.8\sigma$ to $2.9\sigma$. When combined with SNe, the inclusion of the DESI DR1 FS data increases the preference for the dynamical dark energy \resub{by $0.2\sigma$ across all three SNe datasets. The overall preference for evolving dark energy ranges from $2.9\sigma$ to $4.3\sigma$, depending on the choice of supernova dataset.}
\resub{These frequentist tests are consistent with the Bayesian results derived from the marginalized two-dimensional posterior in the $(w_0,w_a)$ plane reported in Tab.~\ref{tab:w0wa}, with the largest difference of $0.6\sigma$ found for the CMB+BAO+DESY5 analysis. In that case, the Bayesian significance reaches $4.7\sigma$, where uncertainties in the normalization of the two-dimensional posterior become non-negligible due to the sparse sampling of the distribution tails in the MCMC analysis. We therefore caution against over-interpreting the Bayesian result in this instance. 
}

\subsection{Massive Neutrinos}\label{sec:mnu}
\begin{figure}[!t]
	\centering
	\includegraphics[width=0.45\textwidth]{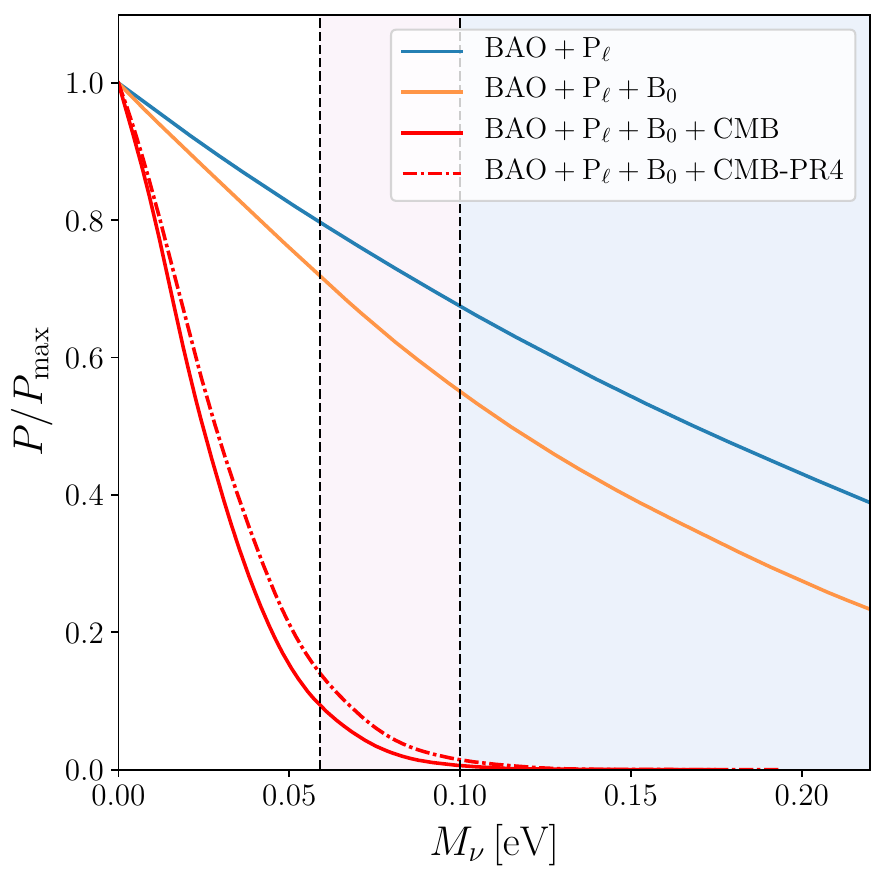}
    \includegraphics[width=0.45\textwidth]{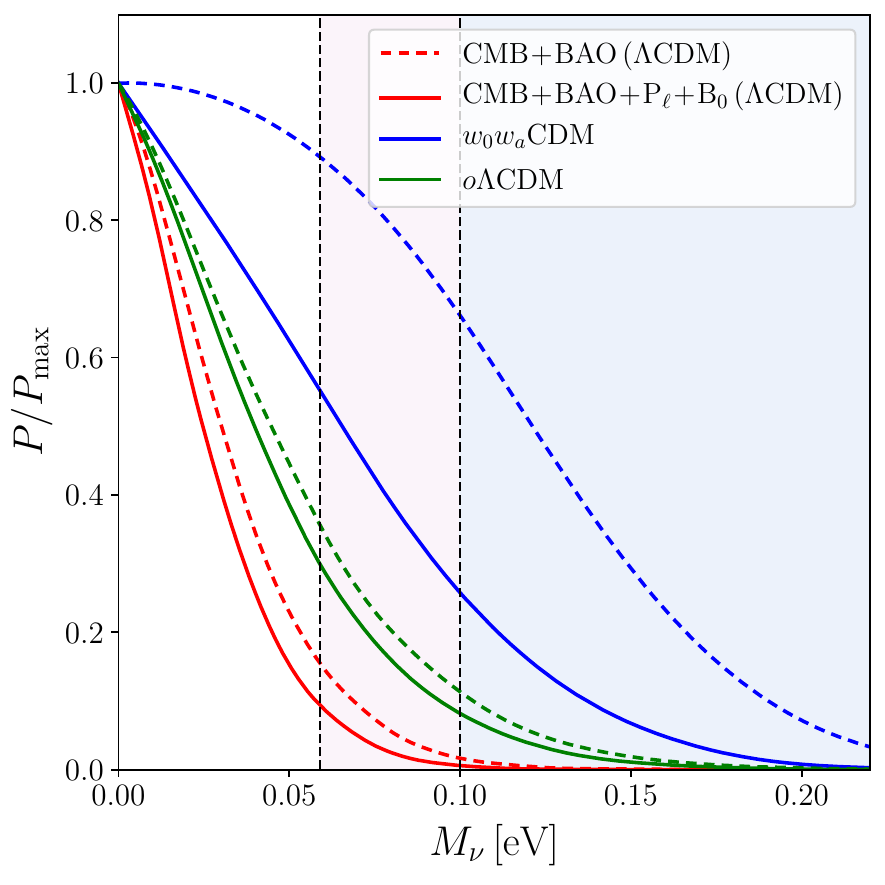}
	\caption{\textbf{Massive neutrinos}: 1D marginalized posterior distributions for the sum of neutrino masses from different datasets. {\it Left panel:} Constraints on $M_\nu$ obtained in a $\ld$ background using the DESI DR2 BAO measurements, further including the DESI DR1 power spectrum $P_\ell$ (blue) and the DESI DR1 bispectrum monopole $B_0$ (orange), together with CMB data either from the official \textsc{plik} data release (red) or from the newer \texttt{HiLLiPoP}+\texttt{LoLLiPoP} likelihoods (dot-dashed).
    {\it Right panel:} constraints on $M_\nu$ from the CMB+BAO (dashed) and $\rm \cmb+\bao+P_\ell+B_0$ (solid) dataset combinations in the $\ld$ (red), $\wa$ (blue) and $\OmK$ (red) backgrounds. 
    The minimal masses for the normal and inverted mass ordering scenarios, corresponding respectively to $M_\nu\geq 0.059\eV$ and $M_\nu\geq 0.10\eV$, are shown by the vertical dashed lines and shaded regions. 
    }
    \label{fig:mnu}
\end{figure}

\noindent 
Lastly, we present constraints on the sum of neutrino masses $M_\nu$. As specified above, we adopt a degenerate mass spectrum and impose a minimal physical prior $M_\nu>0$.
We explore the cosmological implications of massive neutrinos in three different cosmological backgrounds specified by the $\ld$, $\wa$ and $\OmK$ models. 
One-dimensional marginalized constraints are given in Tab.~\ref{fig:mnu} and Fig.~\ref{fig:mnu} shows the one-dimensional posteriors on the sum of neutrino masses in the $\ld$ background (left panel) and in the other cosmological backgrounds (right panel).
Hereafter, we quote 95\% upper limits on the sum of neutrino masses. 

\begin{table*}[!t]
    \centering
    \begin{tabular}{lccccccc}
    \toprule
    Model/Dataset 
    & $M_\nu\,{\rm [eV]}$ 
    & $\Omega_m$ 
    & $H_0$ 
    & $\sigma_8$ 
    & $w_0$ or $10^3\Omk$
    & $w_a$
    & NO/IO
    \\
    \midrule
    $\bm{\Lambda}$\textbf{CDM+}$\bm{M_\nu}$ &  &  &  &  &  & \\
    $\bao+P_\ell$ 
    & $\enspace <0.436\enspace$ 
& $\enspace 0.2941_{-0.0073}^{+0.0073}\enspace$ 
& $\enspace 68.48_{-0.65}^{+0.65}\enspace$ 
& $\enspace 0.826_{-0.033}^{+0.033}\enspace$ 
& $\enspace - \enspace$
& $\enspace - \enspace$
& $0.5\sigma$
    \\
    $\bao+P_\ell+B_0$ 
    & $<0.320$ 
& $0.2977_{-0.0070}^{+0.0070}$ 
& $68.71_{-0.61}^{+0.61}$ 
& $0.813_{-0.030}^{+0.030}$ 
& $-$
& $-$
& $0.7\sigma$
    \\
    $\bao+P_\ell+B_0+\cmb$ 
    & $<0.0592$ 
& $0.2970_{-0.0035}^{+0.0035}$ 
& $68.79_{-0.29}^{+0.29}$ 
& $0.8167_{-0.0056}^{+0.0062}$ 
& $-$
& $-$
& $3.0\sigma$
    \\
    $\bao+P_\ell+B_0+\cmbpr$ 
    & $<0.0669$ 
& $0.2969_{-0.0035}^{+0.0034}$ 
& $68.68_{-0.28}^{+0.28}$ 
& $0.8163_{-0.0049}^{+0.0062}$ 
& $-$
& $-$
& $2.7\sigma$
    \\
    $\cmb+\bao$ 
    & $<0.0687$ 
& $0.2999_{-0.0038}^{+0.0038}$ 
& $68.56_{-0.30}^{+0.30}$ 
& $0.8178_{-0.0058}^{+0.0067}$ 
& $-$
& $-$
& $2.6\sigma$
    \\
    \midrule
    $\bm{w_0w_a}$\textbf{CDM+$\bm{M_\nu}$} &  &  &  &  &  & \\
    $\cmb+\bao$
    & $<0.166$ 
& $0.351_{-0.022}^{+0.023}$ 
& $63.86_{-2.19}^{+1.67}$ 
& $0.783_{-0.019}^{+0.019}$ 
& $-0.43_{-0.21}^{+0.24}$ 
& $-1.72_{-0.69}^{+0.66}$ 
& $1.1\sigma$
    \\
    $\cmb+\bao+P_\ell+B_0$ 
    & $<0.130$ 
& $0.327_{-0.023}^{+0.015}$ 
& $66.01_{-1.68}^{+2.05}$ 
& $0.799_{-0.015}^{+0.018}$ 
& $-0.65_{-0.23}^{+0.16}$ 
& $-1.11_{-0.44}^{+0.66}$ 
& $1.5\sigma$
    \\
    $\cmbpr+\bao+P_\ell+B_0$ 
    & $<0.148$ 
& $0.325_{-0.022}^{+0.015}$ 
& $66.09_{-1.59}^{+2.06}$ 
& $0.799_{-0.014}^{+0.019}$ 
& $-0.68_{-0.23}^{+0.15}$ 
& $-1.02_{-0.42}^{+0.66}$ 
& $1.3\sigma$
    \\
    $\cmb+\bao+P_\ell+B_0+{\rm Pantheon}+$  
    & $<0.104$ 
& $0.3076_{-0.0054}^{+0.0054}$ 
& $67.88_{-0.56}^{+0.56}$ 
& $0.8136_{-0.0085}^{+0.0084}$ 
& $-0.850_{-0.052}^{+0.053}$ 
& $-0.54_{-0.18}^{+0.21}$ 
& $1.9\sigma$
    \\
    $\cmb+\bao+P_\ell+B_0+{\rm Union3}$  
    & $<0.119$ 
& $0.3207_{-0.0089}^{+0.0080}$ 
& $66.56_{-0.82}^{+0.82}$ 
& $0.8035_{-0.0094}^{+0.0103}$ 
& $-0.709_{-0.090}^{+0.082}$ 
& $-0.94_{-0.27}^{+0.31}$ 
& $1.6\sigma$
    \\
    $\cmb+\bao+P_\ell+B_0+{\rm DESY5}$  
    & $<0.116$ 
& $0.3150_{-0.0054}^{+0.0054}$ 
& $67.13_{-0.53}^{+0.53}$ 
& $0.8078_{-0.0086}^{+0.0086}$ 
& $-0.769_{-0.055}^{+0.054}$ 
& $-0.77_{-0.20}^{+0.23}$ 
& $1.7\sigma$
    \\
    \midrule
    $\bm{\OmK}$+$\bm{M_\nu}$ &  &  &  &  &  & \\
    $\cmb+\bao$
    & $<0.103$ 
& $0.3019_{-0.0043}^{+0.0039}$ 
& $68.71_{-0.33}^{+0.33}$ 
& $0.8196_{-0.0067}^{+0.0083}$ 
& $1.9_{-1.3}^{+1.3}$ 
& $-$ 
& $1.9\sigma$
    \\
    $\cmb+\bao+P_\ell+B_0$
    & $<0.0969$ 
& $0.2994_{-0.0039}^{+0.0039}$ 
& $68.95_{-0.31}^{+0.31}$ 
& $0.8191_{-0.0063}^{+0.0077}$ 
& $2.2_{-1.2}^{+1.2}$ 
& $-$ 
& $2.0\sigma$
    \\
    $\cmbpr+\bao+P_\ell+B_0$
    & $<0.114$ 
& $0.2987_{-0.0040}^{+0.0037}$ 
& $68.82_{-0.31}^{+0.32}$ 
& $0.8173_{-0.0059}^{+0.0083}$ 
& $1.8_{-1.3}^{+1.2}$ 
& $-$ 
& $1.7\sigma$
    \\
    \bottomrule
    \end{tabular}
    \caption{\textbf{Massive neutrinos}: Constraints on cosmological parameters where the sum of neutrino masses is allowed to vary assuming a $M_\nu>0$ prior. We quote the mean and 68\% confidence intervals for all parameters, except $M_\nu$ for which 95\% upper limits are provided. The neutrino constraints are visualized in Fig.\,\ref{fig:mnu}.
    \resub{The last column reports the significance for rejecting the inverted mass ordering (IO) in preference for normal ordering (NO), obtained by computing the posterior probability of the IO region, $p(M_\nu>0.098\eV)$, and converting this probability into an equivalent one-dimensional Gaussian $N\sigma$ significance.}
    \label{tab:mnu}
    }
\end{table*}

We begin by considering a $\ld$ background. The DESI FS measurement allows the sum of neutrino masses to be constrained independently of the CMB. The combined analysis of the DESI DR2 BAO and DESI DR1 power spectrum yields  $M_\nu<0.436\eV$; this is tightened by $30\%$ when including the bispectrum monopole, with $M_\nu< 0.320\eV$. The bispectrum contribution helps break the geometric degeneracies present in CMB data, such as $M_\nu-H_0$, resulting in a significant improvement in the $M_\nu$ constraint.\footnote{The DESI DR1 bispectrum is not directly sensitive to the effects of neutrino masses, such as the suppression of power induced by neutrino free-streaming, due to the limited signal-to-noise of the higher-order statistics on large scales. Future DESI data releases may change this situation.} 
To our knowledge, this is the tightest CMB-independent bound on the sum of neutrino masses from cosmological data, \citep[cf.,][]{Philcox:2020vvt,Ivanov:2021zmi}.~\footnote{A tighter bound on the sum of neutrino masses can be obtained by imposing a stronger prior on the spectral index, $n_s$, from \textit{Planck}~\cite{DESI:2024hhd}.}
Our constraint improves upon the public DESI DR1 result, $M_\nu<0.409\eV$~\citep{DESI:2024hhd}, by more than 20\%.
When CMB information is included, the constraint strengthens to $M_\nu<0.0592\eV$, which excludes the inverted mass ordering scenario at the $3.0\sigma$ level. Here, the addition of the DESI DR1 FS information tightens the bound by approximately $15\%$ relative to the $\cmb+\bao$ result. When using the newer $\cmbpr$ data, we find a slightly weaker bound of  $M_\nu<0.0669\eV$, consistent with both the normal and inverted neutrino mass hierarchies within \resub{$1.7\sigma$} and \resub{$2.7\sigma$}, respectively. This is attributed to the smaller lensing power excess preferred by the alternative \texttt{HiLLiPoP}/\texttt{LoLLiPoP} likelihoods 
as compared with \texttt{Plik}~\cite{Tristram:2023haj}.

When allowing the equation-of-state parameters $w_0$ and $w_a$ to vary, we find significantly weaker constraints on the neutrino mass bound (see Fig.\,\ref{fig:mnu}), which becomes consistent with both the normal and inverted mass ordering scenarios at $95\%$ CL for all combinations of datasets.
Here, we find significant utility of the FS data, with the upper limit on $M_\nu$ tightening by more than $20\%$ relative to the $\cmb+\bao$ result.
To further constrain the expansion history in the $\wa$ background, we include the SNe measurements, which give consistent parameter constraints across the three samples, tightening the neutrino mass bound by approximately $20\%$.
\resub{In the $\wa$ background, the data cannot discriminate between the neutrino mass ordering scenarios, with all analyses remaining consistent with the inverted mass ordering scenario within $2\sigma$.}

Allowing spatial curvature to vary also relaxes the upper limit on the neutrino mass, although the effect is smaller than for $\wa$. In this case, the inclusion of the FS information improves the neutrino mass constraint by about $10\%$. We do not include the supernova data here as they do not provide a significant improvement in parameter constraints in the $\OmK$ scenario. \resub{The $\cmb+\bao+P_\ell+B_0$ data disfavor the inverted mass ordering scenario at the $2\sigma$ level, whilst the other analyses remain consistent with both neutrino mass hierarchies.}

Finally, we note that the bounds given above assume the physical prior $M_\nu>0$. The 95\% upper limits would increase considerably when adopting more restrictive priors from terrestrial neutrino oscillation experiments ($M_\nu\geq0.058\eV$ or $M_\nu\geq0.098\eV$), e.g.,~\cite{DESI:2025zgx,DESI:2025ejh}. 
\resub{It will also be interesting to explore the implications of negative `effective' neutrino masses, which have recently been discussed~\cite{Green:2024xbb,Craig:2024tky,Elbers:2024sha}. When extending $M_\nu$ to negative values, most of the posterior mass and its peak lie at $M_{\nu,{\rm eff}}<0$, indicating some tensions between the CMB and DESI results within the $\ld$ model~\cite{DESI:2025ejh}.
We do not expect the results to change significantly after incorporating the full-shape measurements.
We leave this task for future work.}
\resub{Finally, we note that in all DESI analyses involving massive neutrinos, the measurements of the weak-lensing amplitude $S_8$ are consistent with the joint DES Y3+KiDS-1000 analysis~\cite{Kilo-DegreeSurvey:2023gfr} within $2.2\sigma$ and with the more recent KiDS-Legacy constraint~\cite{Wright:2025xka} within $1\sigma$.}

\section{Conclusions}\label{sec:conclusions}
\noindent

In this work, we have presented constraints on non-minimal cosmological models using the full-shape (FS) DESI power spectrum and bispectrum, analyzed in the framework of the EFT for large-scale structure. 
Specifically, we have considered models with non-zero spatial curvature, 
dynamical dark energy, and neutrino masses, in all cases considering the FS data in conjunction with CMB, BAO and SNe datasets. Our analysis has been performed using the independent pipeline built in \paperone, utilizing the public DESI DR1 data release \citep{DESI:2025fxa}. Our pipeline extends the official 
EFT-based measurements by the DESI collaboration in a number of ways, most notably by the addition of the redshift-space galaxy power spectrum hexadecapole and the bispectrum monopole. We additionally employ a new set of priors on EFT parameters, which effectively mitigates parameter projection effects in extended cosmological scenarios. This allows us to obtain the first dark energy measurements from the CMB and combined DESI (BAO+FS) data without relying on supernova distance information (in contrast to e.g.~\cite{DESI:2024hhd}). 

Even without the CMB, we have demonstrated that LSS data can produce strong cosmological constraints. For curvature, we find almost a factor of two improvement found in the $\Omk$ parameter over the BAO-only result, providing strong CMB-independent evidence in favor of a spatially flat Universe. In addition, we have obtained the strongest CMB-independent constraint on the total neutrino mass, made possible due to a substantial improvement from the galaxy bispectrum data.  
When combining with the CMB, we find that the FS likelihood does not appreciably improve the BAO-derived spatial curvature constraints. For both dynamical dark energy and neutrino masses, we find a different story, with the Figure of Merit for the equation of state parameters $w_0,w_a$ increasing by $20\%$ with SNe and by $30\%$ without SNe. The addition of the FS data to the CMB+BAO likelihood shifts the $w_0,w_a$ posteriors towards the cosmological constant values. In combination with SNe likelihoods, we do not find a significant shift, with the shrinking of the posterior volume leading to a slight increase in the evidence for dynamical dark energy by $0.2\sigma$ across all three SNe datasets. Finally, our FS measurements improve the CMB+BAO limits on the neutrino mass both in the minimal $\Lambda$CDM model and with extended expansion histories. In particular, 
we find 95\%CL limits $M_\nu<0.059~\eV$ in $\Lambda$CDM+$M_\nu$, $M_\nu<0.097~\eV$ in $o\Lambda$CDM+$M_\nu$ and $M_\nu<0.13~\eV$ in $w_0w_a$CDM+$M_\nu$, which are
$14\%,~6\%,~22\%,$ stronger than the CMB+BAO only results. 

In future, it will be interesting to explore other extended cosmological scenarios. For instance, \citep{Sailer:2025lxj} proposed an alternative explanation for DESI's expansion-rate discrepancies by invoking a high optical depth $\tau$ (made possible by omitting large-scale polarization information from \textit{Planck}). In addition, it will be important to re-analyse the full combination of BAO, CMB, and FS data
in the context of physical dark energy models that produce effects beyond the phenomenological CPL parameterization, e.g. dark QCD model of~\cite{Khoury:2025txd}.
More generally, our likelihood is expected to improve constraints on cosmological models whose phenomenology goes beyond addressing DESI's results. Such bounds can be readily obtained with our FS power spectrum
and bispectrum likelihood, but are left to future work.

\vskip 8pt
\acknowledgments
{\small
\begingroup
\hypersetup{hidelinks}
\noindent 
We thank Noah Sailer and Maria Tsedrik for insightful discussions. AC acknowledges funding from the Swiss National Science Foundation. The computations in this work were run at facilities supported by the Scientific Computing Core at the Flatiron Institute, a division of the Simons Foundation, as well as at the Helios cluster at the Institute for Advanced Study, Princeton.
OHEP thanks the \href{https://www.flickr.com/photos/198816819@N07/55146242509//}{localls of South Korea} for 
support while this work was being completed.

\noindent This research used data obtained with the Dark Energy Spectroscopic Instrument (DESI). DESI construction and operations is managed by the Lawrence Berkeley National Laboratory. This material is based upon work supported by the U.S. Department of Energy, Office of Science, Office of High-Energy Physics, under Contract No. DE–AC02–05CH11231, and by the National Energy Research Scientific Computing Center, a DOE Office of Science User Facility under the same contract. Additional support for DESI was provided by the U.S. National Science Foundation (NSF), Division of Astronomical Sciences under Contract No. AST-0950945 to the NSF’s National Optical-Infrared Astronomy Research Laboratory; the Science and Technology Facilities Council of the United Kingdom; the Gordon and Betty Moore Foundation; the Heising-Simons Foundation; the French Alternative Energies and Atomic Energy Commission (CEA); the National Council of Humanities, Science and Technology of Mexico (CONAHCYT); the Ministry of Science and Innovation of Spain (MICINN), and by the DESI Member Institutions: \url{www.desi.lbl.gov/collaborating-institutions}. The DESI collaboration is honored to be permitted to conduct scientific research on I’oligam Du’ag (Kitt Peak), a mountain with particular significance to the Tohono O’odham Nation. Any opinions, findings, and conclusions or recommendations expressed in this material are those of the author(s) and do not necessarily reflect the views of the U.S. National Science Foundation, the U.S. Department of Energy, or any of the listed funding agencies.
\endgroup
}

\appendix

\section{Parameter Projection Effects}\label{app:marg}
\noindent
The analysis pipeline used in this work was validated in \paperone in the context of the $\ld$ model. In this work, however, we update the pipeline (as described in Sec.~\ref{sec:data3}) and apply it to beyond-$\ld$ scenarios, thus additional validation is needed. To check for parameter projection effects, we generate a noiseless synthetic data vector using the \textsc{class-pt} pipeline (following \paperone). As a first step, we compute the best-fit $\ld$ model with ($w_0=-1$, $w_a=0$) from the $\cmb+\bao+P_\ell+B_0$ data by fitting the six cosmological parameters $\{\omega_{cdm}, \omega_{b}, H_0, \ln(10^{10}A_s), n_s,\tau\}$, together with six copies of ($b_1\sigma_8$, $b_2\sigma_8^2$, $b_{\mathcal{G}_2}\sigma_8^2$), corresponding to each DESI data chunk in the FS analysis. 
This best-fit model is then used to produce the mock data of the galaxy power spectrum and bispectrum at six different redshifts, simulating the DESI samples, as well as the DESI BAO DR2 and CMB measurements. To produce the CMB mock data, we use the likelihood \textsc{fake\_planck\_realistic} included in the \textsc{Montepython} code~\cite{Audren:2012wb,Brinckmann:2018cvx}. Specifically, we generate mock temperature, polarization and CMB
lensing potential data, along with the corresponding noise spectra, adopting the recommended settings designed to reproduce the characteristics of the full Planck mission~\cite{Aghanim:2018eyx}.\footnote{The resulting CMB mock data only approximately describe the real Planck sensitivity, as they rely on a Gaussian CMB likelihood and neglect correlations between temperature and polarization noise~\cite{Brinckmann:2018owf}; this is not expected to affect our conclusions.}

We fit the synthetic data using the same theoretical model employed in the real data analysis. For the $P_\ell$, $B_0$, and BAO mock data, we adopt the same covariance matrices as in the data analyses, while for the CMB mock data the uncertainties are estimated from the simulated temperature and polarization noise spectra~\cite{Brinckmann:2018owf}. In analyses that exclude the CMB information, we fix $\tau$ to the value used to generate the mock data and impose Gaussian priors on $\omega_b$ and $n_s$, centered on their mock-generation values with widths matching those adopted in the main analyses of this work (see Sec.~\ref{sec:data3}). Since the mock data vector is generated with the same theory pipeline used in the fit, the true theory parameter values exactly minimize the likelihood, such that any shift in the posterior means from these input values quantifies the bias in parameter recovery induced by the Bayesian marginalization effects.

\begin{figure}[!t]
	\centering
	\includegraphics[width=0.8\textwidth]{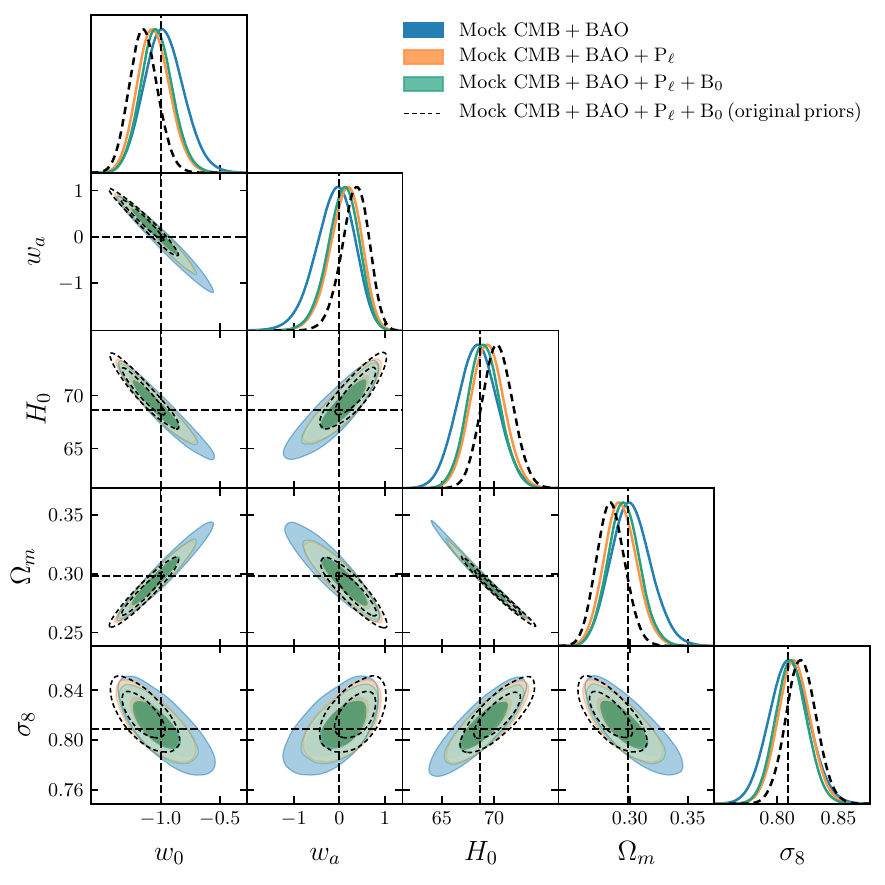}
	\caption{\textbf{Projection Effects in $\bm{w_0w_a}$\textbf{CDM}}: Cosmological parameter constraints estimated from the synthetic mock data generated with our pipeline in the $\wa$ model. Results are presented for three data combinations: $\cmb+\bao$ (blue), $\cmb+\bao+P_\ell$ (orange) and $\cmb+\bao+P_\ell+B_0$ (green). The dashed black contours show the posterior distributions from the $\cmb+\bao+P_\ell+B_0$ analysis derived using the original EFT priors from \paperone.
     }
    \label{fig:synth_wa}
\end{figure}

We begin with the $\wa$ model. Tab.~\ref{tab:synth_wa} presents the impact of parameter projection effects, with the one- and two-dimensional posterior distributions for the selected parameters shown in Fig.~\ref{fig:synth_wa}.
\begin{table}[!t]
    \centering
  \begin{tabular}{|c|cccccccc|} \hline
    \textbf{Mock data} 
    & $w_0$
    & $w_a$
    & $\omega_{\rm cdm}$
    & $H_0$ 
    & ${\ln(10^{10}A_s)}$ 
    & $n_s$
    & $\Omega_m$ 
    & $\sigma_8$
    \\
    \hline
    $\cmb+\bao$
    & $\enspace 0.19\sigma \enspace$ %
    & $\enspace -0.25\sigma \enspace$ %
    & $\enspace 0.21\sigma \enspace$ %
    & $\enspace -0.10\sigma \enspace$ %
    & $\enspace 0.14\sigma \enspace$ %
    & $\enspace -0.12\sigma \enspace$ %
    & $\enspace 0.16\sigma \enspace$ %
    & $\enspace 0.07\sigma \enspace$ %
    \\
    $\cmb+\bao+P_\ell$
    & $-0.41\sigma$ %
    & $0.37\sigma$ %
    & $-0.20\sigma$
    & $0.46\sigma$ 
    & $0.11\sigma$ 
    & $0.16\sigma$ 
    & $-0.44\sigma$ 
    & $0.38\sigma$ 
    \\
    $\cmb+\bao+P_\ell+B_0$
    & $-0.23\sigma$ %
    & $0.21\sigma$ %
    & $-0.17\sigma$
    & $0.27\sigma$ 
    & $0.10\sigma$ 
    & $0.16\sigma$ 
    & $-0.25\sigma$ 
    & $0.21\sigma$
    \\\hline
    $\cmb+\bao+P_\ell+B_0$ (original priors)
    & $-1.19\sigma$ %
    & $1.13\sigma$ %
    & $-0.41\sigma$
    & $1.17\sigma$ 
    & $0.10\sigma$ 
    & $0.34\sigma$ 
    & $-1.19\sigma$ 
    & $0.88\sigma$
    \\
  \hline
    \end{tabular}
    \caption{\textbf{Projection Effects in $\bm{w_0w_a}$\textbf{CDM}}: Differences between the true parameter values and the recovered posterior means from analyses of synthetic mock data in the $\wa$ model. Results are shown for three data combinations using the baseline EFT priors of this work, expressed in units of the standard deviation from the corresponding analysis. We additionally show the results for the $\cmb+\bao+P_\ell+B_0$ analysis when adopting the earlier EFT priors from \paperone (original priors). The updated analysis pipeline mitigates strong parameter projection effects in the $\wa$ scenario. 
    }
\label{tab:synth_wa}
\end{table} 
In the $\cmb+\bao$ analysis, the marginalization bias is minimal, remaining below $0.25\sigma$ for all parameters. When adding the power spectrum, the magnitude of projection effects increases to $\approx 0.4\sigma$ driven by the presence of many nuisance parameters in the FS analysis which open new degeneracy directions; adding the bispectrum reduces these shifts to $\lesssim0.2\sigma$ for all parameters, since this does not contain many new parameters.
If we instead use the original EFT priors from \paperone (without the $A_{\rm AP}$ rescaling), we find significantly stronger projection effects in the $\wa$ scenario, in agreement with previous findings by DESI~\cite{DESI:2024jxi,DESI:2024hhd}.
We conclude that the updated analysis pipeline employed in this work yields robust constraints in the $\wa$ model, owing to the reparameterization of the EFT parameters introduced in Sec.~\ref{sec:data3}. 

\begin{figure}[!t]
	\centering
	\includegraphics[width=0.95\textwidth]{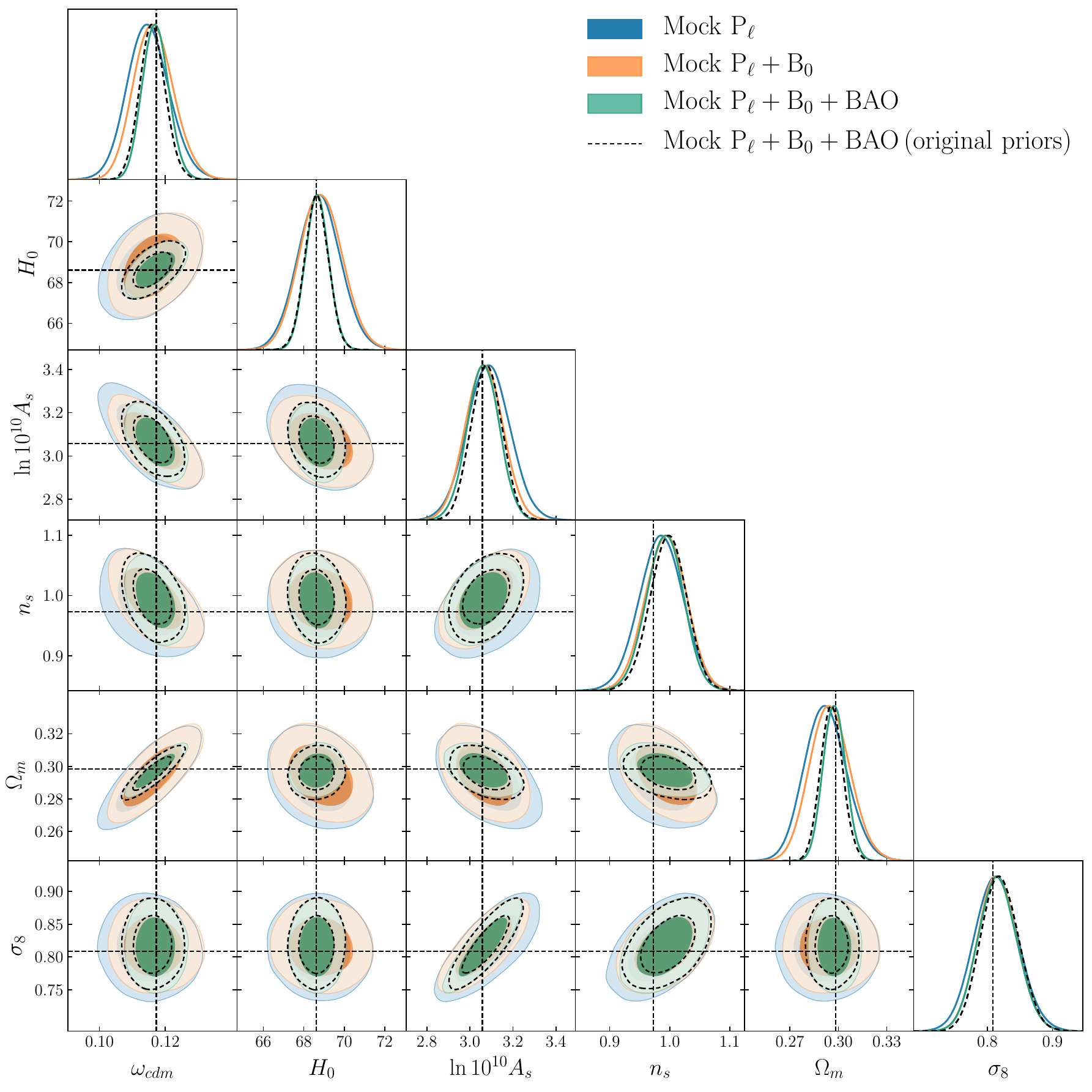}
	\caption{\textbf{Projection Effects in $\bm{\Lambda}$\textbf{CDM}}: As Fig.\,\ref{fig:synth_wa} but for the $\ld$ model. We find good recovery of the input cosmological parameters indicating that prior volume effects are well controlled.
     }
    \label{fig:synth_ld}
\end{figure}

We also revisit the impact of parameter projection effects in the $\ld$ model. Fig.~\ref{fig:synth_ld} shows the one- and two-dimensional marginalized posterior distributions for the cosmological parameters, with the corresponding parameter shifts listed in Tab.~\ref{tab:synth_ld}.
\begin{table}[!t]
    \centering
  \begin{tabular}{|c|cccccc|} \hline
    \textbf{Mock data} 
    & $\omega_{\rm cdm}$
    & $H_0$ 
    & ${\ln(10^{10}A_s)}$ 
    & $n_s$
    & $\Omega_m$ 
    & $\sigma_8$
    \\
    \hline
    $P_\ell$
    & $\enspace -0.37\sigma \enspace$ %
    & $\enspace 0.14\sigma \enspace$ %
    & $\enspace 0.28\sigma \enspace$ %
    & $\enspace 0.35\sigma \enspace$ %
    & $\enspace -0.48\sigma \enspace$ %
    & $\enspace 0.15\sigma \enspace$ %
    \\
    $P_\ell+B_0$
    & $-0.12\sigma$
    & $0.24\sigma$ 
    & $0.09\sigma$ 
    & $0.61\sigma$ 
    & $-0.28\sigma$ 
    & $0.24\sigma$ 
    \\
    $P_\ell+B_0+{\rm BAO}$
    & $-0.06\sigma$
    & $0.07\sigma$ 
    & $0.05\sigma$ 
    & $0.61\sigma$ 
    & $-0.12\sigma$ 
    & $0.23\sigma$
    \\\hline
    $P_\ell+B_0+{\rm BAO}$ (original priors)
    & $-0.27\sigma$
    & $0.04\sigma$ 
    & $0.24\sigma$ 
    & $0.80\sigma$ 
    & $-0.36\sigma$ 
    & $0.36\sigma$
    \\
  \hline
    \end{tabular}
    \caption{\textbf{Projection Effects in $\bm{\Lambda}$\textbf{CDM}}: As Tab.\,\ref{tab:synth_wa} but for the $\ld$ model. The updated analysis pipeline alleviate the marginalization bias in parameter recovery.  
    }
\label{tab:synth_ld}
\end{table} 
We find that using the updated analysis pipeline reduces the marginalization-induced shifts in the DESI-only analyses compared to the earlier pipeline used in \paperone.
In particular, the $P_\ell$-only shifts caused by marginalization effects decrease from $0.9\sigma$ with the original EFT priors to $0.5\sigma$ with the updated pipeline, for all parameters except $n_s$. When adding $B_0$, they decreases from $0.6\sigma$ to $0.3\sigma$, and further reduce to $\approx 0.2\sigma$ when adding BAO information. In this case, we do not combine with the CMB mock data, because the parameter projection effects are already strongly suppressed in the $\ld$ model once CMB information is included~\cite{DESI:2024hhd}.

All in all, our results demonstrate that the updated analysis pipeline is robust and can be applied to DESI data in beyond-$\ld$ scenarios. 

\section{Results with original EFT priors}\label{app:old}
\noindent
In this work, we update the analysis pipeline with respect to \paperone as described in Sec.~\ref{sec:data3}. In this Appendix, we present parameter constraints obtained using the original EFT priors from \paperone in order to elucidate the prior dependence of cosmological results. Throughout this section, we refer to the priors from \paperone as the `original priors', for simplicity.
The parameter constraints for models with free spatial curvature, dynamical dark energy, and massive neutrinos are given in Tabs.~\ref{tab:OmK_old}, \ref{tab:w0wa_old} and~\ref{tab:mnu_old}, respectively. For the massive neutrino case, we provide results for three cosmological backgrounds --  $\ld$, $\wa$ and $\OmK$ models -- as in the main analysis of this work. 
\begin{table*}[!t]
    \centering
    \begin{tabular}{lcccc}
    \toprule
    Dataset 
    & $10^3\Omk$ 
    & $\Omega_m$ 
    & $H_0$ 
    & $\sigma_8$  \\
    \midrule
    $\bao+P_\ell$ 
    & $0.5_{-23.4}^{+23.2}$ 
& $0.2920_{-0.0074}^{+0.0074}$ 
& $68.69_{-1.06}^{+1.07}$ 
& $0.832_{-0.034}^{+0.034}$ 
    \\
    $\bao+P_\ell+B_0$ 
    & $-8.9_{-21.1}^{+21.3}$ 
& $0.2966_{-0.0068}^{+0.0068}$ 
& $69.16_{-0.99}^{+0.99}$ 
& $0.817_{-0.031}^{+0.031}$ 
    \\\hline
    $\cmb+P_\ell$ 
    & $4.4_{-1.8}^{+1.9}$ 
& $0.2936_{-0.0068}^{+0.0068}$ 
& $69.74_{-0.77}^{+0.77}$ 
& $0.8185_{-0.0064}^{+0.0064}$ 
    \\
    $\cmb+P_\ell+B_0$ 
    & $4.4_{-1.8}^{+1.9}$ 
& $0.2946_{-0.0067}^{+0.0067}$ 
& $69.67_{-0.76}^{+0.76}$ 
& $0.8179_{-0.0062}^{+0.0062}$ 
    \\
    $\cmb+P_\ell+B_0+\bao$ 
    & $2.6_{-1.1}^{+1.1}$ 
& $0.3005_{-0.0036}^{+0.0036}$ 
& $68.89_{-0.30}^{+0.30}$ 
& $0.8149_{-0.0057}^{+0.0058}$ 
    \\
    $\cmbpr+P_\ell+B_0+\bao$ 
    & $2.2_{-1.1}^{+1.1}$ 
& $0.2995_{-0.0035}^{+0.0035}$ 
& $68.78_{-0.30}^{+0.30}$ 
& $0.8138_{-0.0052}^{+0.0051}$ 
    \\
    \bottomrule
    \end{tabular}
    \caption{\textbf{Spatial curvature (original priors)}: Mean and 68\% confidence intervals on cosmological parameters in the $\OmK$ analyses, obtained using the original EFT priors from \paperone. 
    }
    \label{tab:OmK_old}
\end{table*}
\begin{table*}[!t]
    \centering
    \begin{tabular}{lccccc}
    \toprule
    Dataset 
    & $w_0$ 
    & $w_a$ 
    & $\Omega_m$ 
    & $H_0$ 
    & $\sigma_8$  \\
    \midrule
    $\cmb+\bao+P_\ell$ 
    & $\enspace -0.77_{-0.18}^{+0.14}\enspace$ 
& $\enspace -0.85_{-0.38}^{+0.53}\enspace$ 
& $\enspace 0.313_{-0.018}^{+0.013}\enspace$ 
& $\enspace 67.45_{-1.50}^{+1.76}\enspace$ 
& $\enspace 0.810_{-0.013}^{+0.014}\enspace$ 
    \\
    $\cmb+\bao+P_\ell+B_0$ 
    & $-0.75_{-0.18}^{+0.12}$ 
& $-0.85_{-0.35}^{+0.51}$ 
& $0.316_{-0.017}^{+0.012}$ 
& $67.10_{-1.31}^{+1.67}$ 
& $0.806_{-0.012}^{+0.013}$ 
    \\
    $\cmb+\bao+P_\ell+B_0+{\rm Pantheon}+$  
    & $-0.850_{-0.051}^{+0.051}$ 
& $-0.58_{-0.18}^{+0.20}$ 
& $0.3073_{-0.0052}^{+0.0052}$ 
& $67.94_{-0.54}^{+0.55}$ 
& $0.8115_{-0.0076}^{+0.0076}$ 
    \\
    $\cmb+\bao+P_\ell+B_0+{\rm Union3}$  
    & $-0.725_{-0.088}^{+0.080}$ 
& $-0.93_{-0.25}^{+0.29}$ 
& $0.3185_{-0.0085}^{+0.0076}$ 
& $66.80_{-0.80}^{+0.80}$ 
& $0.8039_{-0.0085}^{+0.0085}$ 
    \\
    $\cmb+\bao+P_\ell+B_0+{\rm DESY5}$  
    & $-0.769_{-0.054}^{+0.054}$ 
& $-0.81_{-0.19}^{+0.22}$ 
& $0.3144_{-0.0053}^{+0.0052}$ 
& $67.21_{-0.52}^{+0.53}$ 
& $0.8067_{-0.0073}^{+0.0073}$ 
    \\
    \bottomrule
    \end{tabular}
    \caption{\textbf{Dynamical Dark Energy (original priors)}: Mean and 68\% confidence intervals on cosmological parameters in the $\wa$ analyses, derived using the original EFT priors from \paperone. 
    }
    \label{tab:w0wa_old}
\end{table*}
\begin{table*}[!t]
    \centering
    \begin{tabular}{lcccccc}
    \toprule
    Model/Dataset 
    & $M_\nu\,{\rm [eV]}$ 
    & $\Omega_m$ 
    & $H_0$ 
    & $\sigma_8$ 
    & $w_0$ or $\Omk$
    & $w_a$
    \\
    \midrule
    $\bm{\Lambda}$\textbf{CDM+}$\bm{M_\nu}$ &  &  &  &  &  & \\
    $\bao+P_\ell$ 
    & $\enspace <0.435\enspace$ 
& $\enspace 0.2934_{-0.0073}^{+0.0073}\enspace$ 
& $\enspace 68.48_{-0.65}^{+0.65}\enspace$ 
& $\enspace 0.829_{-0.032}^{+0.032}\enspace$ 
& $\enspace - \enspace$
& $\enspace - \enspace$
    \\
    $\bao+P_\ell+B_0$ 
    & $<0.311$ 
& $0.2969_{-0.0070}^{+0.0070}$ 
& $68.71_{-0.61}^{+0.61}$ 
& $0.819_{-0.030}^{+0.030}$ 
& $-$
& $-$
    \\
    $\bao+P_\ell+B_0+\cmb$ 
    & $<0.0585$ 
& $0.2968_{-0.0035}^{+0.0035}$ 
& $68.80_{-0.29}^{+0.29}$ 
& $0.8169_{-0.0056}^{+0.0061}$ 
& $-$
& $-$
    \\
    $\bao+P_\ell+B_0+\cmbpr$ 
    & $<0.0657$ 
& $0.2968_{-0.0035}^{+0.0034}$ 
& $68.69_{-0.28}^{+0.28}$ 
& $0.8165_{-0.0049}^{+0.0062}$ 
& $-$
& $-$
    \\
    \midrule
    $\bm{w_0w_a}$\textbf{CDM+$\bm{M_\nu}$} &  &  &  &  &  & \\
    $\cmb+\bao+P_\ell+B_0$ 
    & $<0.116$ 
& $0.314_{-0.017}^{+0.012}$ 
& $67.27_{-1.31}^{+1.64}$ 
& $0.809_{-0.012}^{+0.015}$ 
& $-0.78_{-0.17}^{+0.12}$ 
& $-0.77_{-0.35}^{+0.52}$ 
    \\
    $\cmbpr+\bao+P_\ell+B_0$ 
    & $<0.131$ 
& $0.312_{-0.016}^{+0.011}$ 
& $67.35_{-1.22}^{+1.58}$ 
& $0.809_{-0.012}^{+0.015}$ 
& $-0.80_{-0.17}^{+0.12}$ 
& $-0.67_{-0.33}^{+0.49}$ 
    \\
    $\cmb+\bao+P_\ell+B_0+{\rm Pantheon}+$  
    & $<0.104$ 
& $0.3067_{-0.0053}^{+0.0053}$ 
& $67.98_{-0.55}^{+0.55}$ 
& $0.8146_{-0.0084}^{+0.0085}$ 
& $-0.858_{-0.052}^{+0.052}$ 
& $-0.52_{-0.18}^{+0.21}$ 
    \\
    $\cmb+\bao+P_\ell+B_0+{\rm Union3}$  
    & $<0.115$ 
& $0.3180_{-0.0085}^{+0.0076}$ 
& $66.84_{-0.79}^{+0.79}$ 
& $0.8062_{-0.0095}^{+0.0096}$ 
& $-0.734_{-0.088}^{+0.079}$ 
& $-0.88_{-0.25}^{+0.31}$ 
    \\
    $\cmb+\bao+P_\ell+B_0+{\rm DESY5}$  
    & $<0.114$ 
& $0.3140_{-0.0053}^{+0.0054}$ 
& $67.24_{-0.53}^{+0.52}$ 
& $0.8092_{-0.0085}^{+0.0085}$ 
& $-0.776_{-0.055}^{+0.055}$ 
& $-0.76_{-0.20}^{+0.23}$ 
    \\
    \midrule
    $\bm{\OmK}$+$\bm{M_\nu}$ &  &  &  &  &  & \\
    $\cmb+\bao+P_\ell+B_0$
    & $<0.0960$ 
& $0.2992_{-0.0039}^{+0.0039}$ 
& $68.97_{-0.32}^{+0.32}$ 
& $0.8190_{-0.0063}^{+0.0076}$ 
& $2.2_{-1.2}^{+1.2}$
& $-$ 
    \\
    \bottomrule
    \end{tabular}
    \caption{\textbf{Massive neutrinos (original priors)}: Constraints on cosmological parameters where the sum of neutrino masses is allowed to vary assuming a $M_\nu>0$ prior, computed using the original priors from \paperone. We quote the mean and 68\% confidence intervals for all parameters, except $M_\nu$ for which 95\% upper limits are provided. 
    \label{tab:mnu_old}
    }
\end{table*}

In the $\OmK$ model, the parameter constraints are robust against changes in the EFT priors. In the $\wa$ scenario, we find moderate shifts in the inferred cosmological parameters when SNe are excluded, remaining below $0.6\sigma$ in the $\cmb+\bao+P_\ell$ analysis or $0.7\sigma$ when the bispectrum is added. Interestingly, using the original priors from \paperone yields $w_0$ and $w_a$ posteriors that are more consistent with the cosmological constant. When SNe are included, the effects of parameter projection are suppressed and the results become highly stable with respect to the prior choice, with all parameter shifts below $0.1\sigma$ -- this matches the conclusions of \cite{DESI:2024hhd}.

Finally, we investigate the impact of the EFT priors when allowing the neutrino mass parameter to vary. In the $\ld+M_\nu$ model, the cosmological constraints remain stable with respect to the choice of EFT priors. When the equation-of-state parameters $w_0$ and $w_a$ are allowed to vary, using the original EFT priors leads to the $\approx 10\%$ tighter upper limits on the neutrino mass in the CMB(-PR4)+BAO+$P_\ell+B_0$ analyses, though, as in the fixed-$M_\nu$ case, the neutrino mass constraint becomes largely insensitive to the prior choice when SNe are added. When spatial curvature is allowed to vary, the resulting parameter constraints remain highly robust against changes in the EFT priors.

\bibliographystyle{apsrev4-1}
\bibliography{refs}

\end{document}